\documentclass[aps,reprint,twocolumn,superscriptaddress,citeautoscript,showpacs,amsmath,amssymb,floatfix,prb]{revtex4-1}

\usepackage{amsmath}
\usepackage{amssymb}
\usepackage{graphicx}
\usepackage{times}
\usepackage{bm}
\usepackage{color}
\usepackage{gensymb}
\usepackage{upgreek}
\usepackage{dcolumn}

\newcolumntype{d}[1]{D{.}{.}{#1}}

\begin{document}

\title{Effects of a magnetic field on the fragile antiferromagnetism in YbBiPt}

\author{B.~G.~Ueland}
\affiliation{Ames Laboratory, U.S. DOE, Iowa State University, Ames, Iowa 50011, USA}
\affiliation{Department of Physics and Astronomy, Iowa State University, Ames, Iowa 50011, USA}

\author{A.~Kreyssig}
\affiliation{Ames Laboratory, U.S. DOE, Iowa State University, Ames, Iowa 50011, USA}
\affiliation{Department of Physics and Astronomy, Iowa State University, Ames, Iowa 50011, USA}

\author{E.~D.~Mun}
\affiliation{Department of Physics, Simon Fraser University, Burnaby, British Columbia, Canada V5A 1S6}
\affiliation{Ames Laboratory, U.S. DOE, Iowa State University, Ames, Iowa 50011, USA}
\affiliation{Department of Physics and Astronomy, Iowa State University, Ames, Iowa 50011, USA}

\author{J.~W.~Lynn}
\affiliation{NIST Center for Neutron Research, National Institute of Standards and Technology, Gaithersburg, Maryland 20899, USA}

\author{L.~W.~Harriger}
\affiliation{NIST Center for Neutron Research, National Institute of Standards and Technology, Gaithersburg, Maryland 20899, USA}

\author{D.~K.~Pratt}
\affiliation{NIST Center for Neutron Research, National Institute of Standards and Technology, Gaithersburg, Maryland 20899, USA}

\author{K.~Proke\v{s}}
\affiliation{Helmholtz-Zentrum Berlin f{\"{u}}r Materialien und Energie, Hahn-Meitner-Platz 1, 14109 Berlin, Germany}

\author{Z. H\"{u}sges}
\affiliation{Helmholtz-Zentrum Berlin f{\"{u}}r Materialien und Energie, Hahn-Meitner-Platz 1, 14109 Berlin, Germany}

\author{R.~Toft-Petersen}
\affiliation{Helmholtz-Zentrum Berlin f{\"{u}}r Materialien und Energie, Hahn-Meitner-Platz 1, 14109 Berlin, Germany}

\author{S.~Sauerbrei}
\affiliation{Ames Laboratory, U.S. DOE, Iowa State University, Ames, Iowa 50011, USA}
\affiliation{Department of Physics and Astronomy, Iowa State University, Ames, Iowa 50011, USA}

\author{S.~M.~Saunders}
\affiliation{Ames Laboratory, U.S. DOE, Iowa State University, Ames, Iowa 50011, USA}
\affiliation{Department of Physics and Astronomy, Iowa State University, Ames, Iowa 50011, USA}

\author{Y.~Furukawa}
\affiliation{Ames Laboratory, U.S. DOE, Iowa State University, Ames, Iowa 50011, USA}
\affiliation{Department of Physics and Astronomy, Iowa State University, Ames, Iowa 50011, USA}

\author{S.~L.~Bud'ko}
\affiliation{Ames Laboratory, U.S. DOE, Iowa State University, Ames, Iowa 50011, USA}
\affiliation{Department of Physics and Astronomy, Iowa State University, Ames, Iowa 50011, USA}

\author{R.~J.~McQueeney}
\affiliation{Ames Laboratory, U.S. DOE, Iowa State University, Ames, Iowa 50011, USA}
\affiliation{Department of Physics and Astronomy, Iowa State University, Ames, Iowa 50011, USA}

\author{P.~C.~Canfield}
\affiliation{Ames Laboratory, U.S. DOE, Iowa State University, Ames, Iowa 50011, USA}
\affiliation{Department of Physics and Astronomy, Iowa State University, Ames, Iowa 50011, USA}

\author{A.~I.~Goldman}
\affiliation{Ames Laboratory, U.S. DOE, Iowa State University, Ames, Iowa 50011, USA}
\affiliation{Department of Physics and Astronomy, Iowa State University, Ames, Iowa 50011, USA}

\date{\today}

\begin{abstract}
We present neutron diffraction data for the cubic-heavy-fermion YbBiPt that show broad magnetic diffraction peaks due to the fragile short-range antiferromagnetic (AFM) order persist under an applied magnetic-field $\mathbf{H}$.  Our results for  $\mathbf{H}\perp[\bar{1}~1~0]$ and a temperature of $T=0.14(1)$~K show that the $(\frac{1}{2},\frac{1}{2},\frac{3}{2})$ magnetic diffraction peak can be described by the same two-peak lineshape found for $\mu_{0}H=0$~T below the N\'{e}el temperature of $T_{\text{N}}=0.4$~K.  Both components of the peak exist for $\mu_{0}H\alt1.4$~T, which is well past the AFM phase boundary determined from our new resistivity data. Using neutron diffraction data taken at $T=0.13(2)$~K for $\mathbf{H}\parallel[0~0~1]$ or $[1~1~0]$, we show that domains of short-range AFM order change size throughout the previously determined AFM and non-Fermi liquid regions of the phase diagram, and that the appearance of a magnetic diffraction peak at  $(\frac{1}{2},\frac{1}{2},\frac{1}{2})$ at $\mu_{0}H\approx0.4$~T signals canting of the ordered magnetic moment away from $[1~1~1]$.  The continued broadness of the magnetic diffraction peaks under a magnetic field and their persistence across the AFM phase boundary established by detailed transport and thermodynamic experiments present an interesting quandary concerning the nature of YbBiPt's electronic ground state.
\end{abstract}

\maketitle

\section{Introduction}
Fragile magnetism is associated with strongly-correlated low-temperature electronic states that are highly tunable by parameters such as strain, applied pressure, and magnetic field \cite{Canfield_2016}.  As the temperature $T\rightarrow0$~K, fragile magnetism may result in fluctuations between states separated by very small energy differences and even lead to a quantum-critical point (QCP) and the emergence of quantum-critical fluctuations \cite{Si_2010,Stockert_2011,Stewart_2001,Hertz_1976,Millis_1993}.  In this work, we present results from neutron diffraction and electrical resistivity experiments that detail the response of the fragile antiferromagnetism (AFM) in the heavy-fermion compound YbBiPt to the application of a magnetic field $\mathbf{H}$ strong enough to drive it through the magnetic phase boundary associated with its purported field-induced QCP at $\mu_{0}H=0.4$~T \cite{Mun_2013}.

YbBiPt is a face-centered-cubic (FCC) compound with a remarkably large low-temperature Sommerfeld coefficient of $\gamma\approx8$~J/mol-K$^2$ and is the metallic end member of the $R$BiPt, $R=$~rare earth, series \cite{Mun_2013,Canfield_1991}.  The characteristic temperatures related to the compound's magnetism are all small and comparable: the Kondo temperature is $T_{\text{K}} \approx1$~K \cite{Mun_2013,Fisk_1991}, the Weiss temperature is $\theta_{\text{W}} \approx-2$~K \cite{Mun_2013}, the crystalline-electric field splitting is on the order of $1$ to $10$~K \cite{Robinson_1995, Ueland_2015}, and the N\'{e}el temperature is $T_{\text{N}}=0.4$~K \cite{Mun_2013}.  In light of the compound's FCC lattice and characterization as a low-carrier concentration semimetal, its extremely large $\gamma$ has been proposed to result from magnetic frustration, a low $T_{\text{K}}$,  or a combination of both \cite{Fisk_1991, Hundley_1997}.

Data from transport experiments show a jump in the compound's resistivity at $T_{\text{N}}$ at ambient pressure and $\mu_{0}H=0$~T which is consistent with spin-density-wave type AFM ordering  \cite{Movshovich_1994, Mun_2013}.  A distinct peak also occurs in the heat capacity at $T_{\text{N}}$, and signatures of an AFM transition are seen in other thermodynamic and electrical-transport data as well \cite{Canfield_1991,Fisk_1991,Movshovich_1994,Canfield_1994, Mun_2013, Lacerda_1993, Movshovich_1994b}.  The magnetic phase diagram constructed from such data for $\mathbf{H}$ applied parallel to the $[0~0~1]$ crystalline direction shows that the magnetic field drives $T_{\text{N}} \rightarrow 0$~K at a critical value of $\mu_{0}H_{\text{c}}\approx 0.4$~T.  This point separates the AFM phase from a region characterized by non-Fermi-liquid (nFL) type electrical transport, and increasing the field past $\mu_{0}H\approx0.7$~T results in a crossover to Fermi-liquid (FL) type electrical transport \cite{Mun_2013}. The fragile magnetism of YbBiPt is evidenced by the experimentally observed features corresponding to $T_{\text{N}}$ being extremely sensitive to magnetic field, pressure, and strain \cite{Mun_2013,Lacerda_1993,Movshovich_1994,Movshovich_1994b}, as well as the presence of very broad magnetic neutron diffraction peaks below $T_{\text{N}}$ \cite{Ueland_2014}.

Neutron diffraction experiments for $\mu_{0}H=0$~T and $T<T_{\text{N}}$ found magnetic diffraction peaks corresponding to an AFM propagation vector of $\bm{\tau}=(\frac{1}{2},\frac{1}{2},\frac{1}{2})$ and an ordered-magnetic moment $\bm{\mu}$ lying parallel to $\bm{\tau}$ \cite{Ueland_2014}.  Surprisingly, the peaks' lineshapes are complicated, consisting of two components: a narrow-Gaussian peak that appears below $T_{\text{N}}$ and a broad-Gaussian peak that occurs below $T^{\text{*}}=0.7$~K\cite{Ueland_2014}.  The total integrated intensity of the peak corresponds to $\mu\approx0.8~\mu_{\text{B}}$, however, the ratio of the integrated intensity of the broad component to that of the narrow one is $\approx12$:$1$.  Since the narrow component appears below $T_{\text{N}}$, its associated value of $\mu$ agrees with previous reports that estimate $\mu=0.1$ to $0.25~\mu_{\text{B}}$ \cite{Robinson_1994,Amato_1992}.  The narrow and broad components have corresponding magnetic-correlation lengths of $\xi_{\text{N}}\approx80$~\AA\ and $\xi_{\text{B}} \approx 20$~{\AA}, respectively \cite{Ueland_2014}, which are both smaller than expected for long-range AFM order.

The presence of significant structural disorder that would limit the AFM correlation length has been ruled out by the existence of resolution-limited structural Bragg peaks in neutron diffraction data and by clean high-energy x-ray diffraction patterns \cite{Ueland_2014,Ueland_2015}.  The presence of quantum oscillations in resistance data for $\mu_{0}H\agt6$~T at low temperature also evidences high-quality crystals \cite{Mun_2015}.  The magnetic diffraction peaks are elastic within an energy-resolution window of $\Delta E=0.09$~meV, which means that within such energy resolution they correspond to static AFM order\cite{Ueland_2014}.

In this report, we present results from neutron diffraction experiments performed at $T\le0.75$~K for applied magnetic fields strong enough to traverse the previously identified AFM and FL boundaries \cite{Mun_2013} .  Data for $\mathbf{H}\parallel[\bar{1}~1~0]$ show that the $(\frac{1}{2},\frac{1}{2},\frac{3}{2})$ magnetic diffraction peak remains broad with increasing $H$ and that its height and full width at half maximum (FWHM) smoothly change.  The peak's narrow component exists up to $\mu_{0}H\approx1.2$~T, whereas its broad component persists up to at least $1.6$~T. This indicates that short-range AFM order is not limited to the AFM region.  No sharp changes to the diffraction peak's lineshape occur at the AFM and FL boundaries determined by resistivity.  Rather, depending on the direction of $\mathbf{H}$, the height of the diffraction peak either reaches a maximum near the FL boundary or monotonically decreases upon crossing the AFM boundary.  We argue that the field dependence of the magnetic diffraction peak predominately reflects changes to the populations of domains of short-range AFM order for $\mu_{0}H\alt0.55$~T.

We further report that for $\mathbf{H}\parallel[\bar{1}~1~0]$ a magnetic diffraction peak appears between $\mu_{0}H=0.2$ and $0.4$~T at $T=0.14$~K at the $(\frac{1}{2},\frac{1}{2},\frac{1}{2})$ reciprocal-lattice position.  Its appearance indicates that $\bm{\mu}$ reorients towards $\mathbf{H}$ for $0.4\alt\mu_{0}H\alt1.2$~T, and we calculate the angle by which $\bm{\mu}$ rotates as a function $H$.  We discuss the implications of magnetic domain growth and the reorientation of $\bm{\mu}$ with increasing field using a phase diagram for $\mathbf{H}\parallel[\bar{1}~1~0]$ that is based on new resistivity data.

\section{Experiment}
YbBiPt's unit cell may be described using space group $F\bar{4}3m$ with a room-temperature lattice parameter of $a=6.5953(1)$~\AA \cite{Robinson_1994}.  The nominally $J=\frac{7}{2}$ Yb$^{3+}$ cations sit at the $4d$ Wyckoff positions and should experience a tetrahedral crystalline-electric field \cite{Robinson_1994,Robinson_1995,Ueland_2015}.  Our experiments used single crystals grown out of Bi flux as described previously \cite{Mun_2013, Canfield_1991, Canfield_1992}. 

Standard four-probe ac-resistivity $\rho$ measurements were made in an Oxford dilution refrigerator by applying a periodically oscillating current \textbf{I} with a frequency of $16$~Hz and recording the resulting voltage along \textbf{I}. Pt wires were used as leads and attached to the sample with Epotek H$20$E silver epoxy.  $\mathbf{H}$ was applied along either the $[0~0~1]$, $[\bar{1}~1~0]$, or $[1~1~1]$ crystalline direction, and  \textbf{I} was always applied perpendicular to $\mathbf{H}$.  More details concerning the experimental conditions are given in Ref.~\onlinecite{Mun_2013}. Since the size of the jump in $\rho(T)$ at $T_{\text{N}}$ is very sensitive to the sample preparation and mounting conditions \cite{Mun_2013}, a total of $24$ samples were first screened with $\mu_{0}H=0$~T.  Out of these, a few samples showing the sharpest anomalies at $T_{\text{N}}$ were selected for measurement while applying a magnetic field.

Neutron scattering experiments used several samples consisting of either one single crystal or two coaligned single crystals.  The samples had total masses of $1$ to $3$~g and total mosaic spreads of $\approx 1$\degree\,FWHM.  Given the strong sensitivity of the compound to pressure and strain \cite{Movshovich_1994,Mun_2013,Lacerda_1993}, several methods and glues (CYTOP or HBM X$60$) were used to fix the crystals to a Cu sample holder which was thermally anchored to the mixing chamber of a dilution refrigerator. Cu wire was loosely wrapped around the crystals in order to ensure mechanical stability and provide another thermal path.

Neutron diffraction experiments were performed on the SPINS cold-neutron triple-axis spectrometer at the NIST Center for Neutron Research, and the E-$4$ two-axis diffractometer \cite{Prokes_2017} and FLEXX cold-neutron triple-axis spectrometer \cite{Le_2013} at the Helmholtz-Zentrum Berlin.  Measurements were made with the $(h,h,l)$ reciprocal-lattice plane coincident with the scattering plane. 

Experiments on SPINS utilized a vertically-focused pyrolitic-graphite (PG) monochromator to select incident neutrons with wavelengths of $\lambda=5.504$~\AA, and cooled Be filters were inserted in both the incident and scattered beams to suppress higher-order neutron wavelengths. The neutron guide prior to the monochromator gave an effective collimation of $53^{\prime }$, a $80^{\prime}$ S\"{o}ller-slit collimator was placed between the monochromator and sample, and a radial collimator was inserted after the sample.  A horizontally-focusing PG analyzer selected $\lambda=5.504$~\AA\ neutrons and focused the diffracted beam to a single $^3$He tube detector.   The energy resolution was determined by measuring the incoherent scattering peak of a plastic cylinder: a neutron energy transfer $E$ scan was performed across $E=0$~meV, and the FWHM of the resulting peak gave a value for the resolution of $\Delta E\approx90~\mu\text{eV}$.

Measurements made on FLEXX used a vertically- and horizontally-focused PG monochromator to form a $\lambda=5.464$~\AA\ neutron beam. A velocity selector prior to the monochromator eliminated higher-order wavelength contamination.    No collimators were used, as the effective collimation of the neutron optics was sufficient.  A PG analyzer selected $\lambda=5.464$~\AA\ neutrons and was horizontally focused to a $^3$He tube detector.  The energy resolution was found from the incoherent scattering from a Vanadium rod to be  $\Delta E\approx70~\mu\text{eV}$ by the same procedure used for SPINS.

Experiments on E-$4$ used a vertically-focused PG monochromator that selected $\lambda=2.451$~\AA\ neutrons and a PG filter was placed in the incident beam to reduce higher-order wavelength contamination.  A 40$^{\prime}$  S\"{o}ller slit collimator was inserted between the monochromator and sample and a radial collimator was inserted after the sample.  A $2$-D position-sensitive detector recorded the diffracted neutrons.

The \textsc{dave} \cite{DAVE}, \textsc{lamp} \cite{LAMP}, and \textsc{spectra} \cite{SPECTRA} software packages as well as in-house developed software were used for data reduction and analysis.  Error bars and stated values of uncertainties represent $1$ standard deviation.  Coordinates in reciprocal space are given in reciprocal-lattice units (r.l.u.), where $1$~r.l.u.~$=2\pi/a$.  $\mathbf{Q}$ corresponds to neutron momentum transfer and is given in r.l.u.\ unless otherwise indicated.

\section{Results}
\subsection{Resistivity for $\mathbf{H}\bm{\parallel[0~0~1]}$, $\bm{[\bar{1}~1~0]}$, and $\bm{[1~1~1]}$} \label{sub3_A}
\begin{figure}
	\centering
	\includegraphics[width=1.0\linewidth]{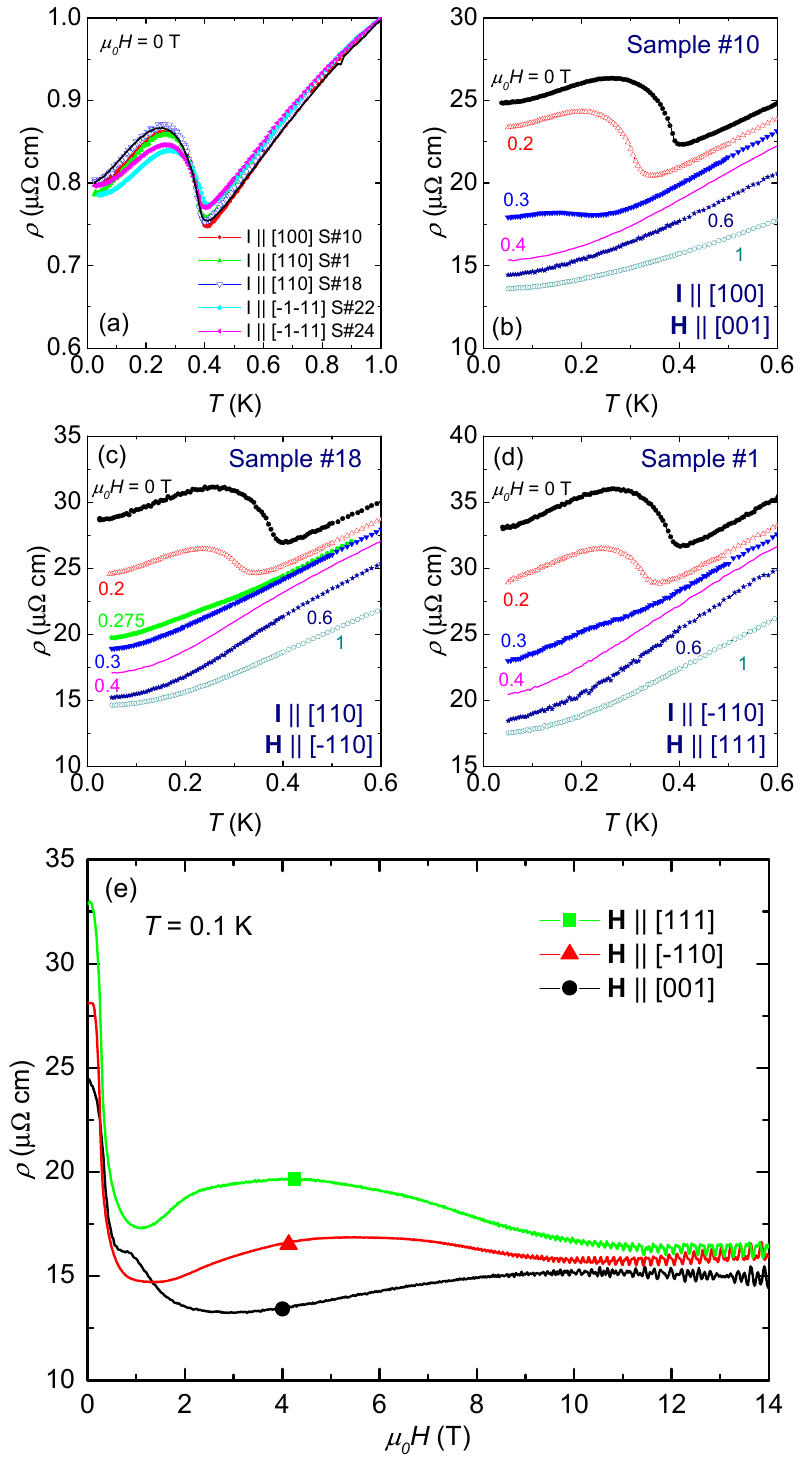}
	\caption{(a) Temperature-dependent electrical resistivity $\rho(T)$ of YbBiPt for a current \textbf{I} with a frequency of $16$~Hz applied along $[1~0~0]$, $[1~1~0]$, or $[\bar{1}~\bar{1}~1]$.  The curves are normalized to $1$ at $T=1$~K, and the legend indicates the direction of \textbf{I} and the sample number corresponding to each plot.  Data for \textbf{I}~$\parallel[0~0~1]$ are from Ref.~\onlinecite{Mun_2013}.  (b)-(d) $\rho(T)$ curves for various values of magnetic field applied along the $[0~0~1]$ (b), $[\bar{1}~1~0]$ (c), or $[1~1~1]$ (d) directions with $\mathbf{I}$ applied ~$\perp\mathbf{H}$. (e) Transverse magnetoresistivity ($\mathbf{I}\perp\mathbf{H}$) at $T=0.1$~K for $\mathbf{H}\parallel[0~0~1]$, $[\bar{1}~0~0]$, or $[1~1~1]$}
	\label{Fig1}
\end{figure}

\begin{figure}
	\centering
	\includegraphics[width=1.0\linewidth]{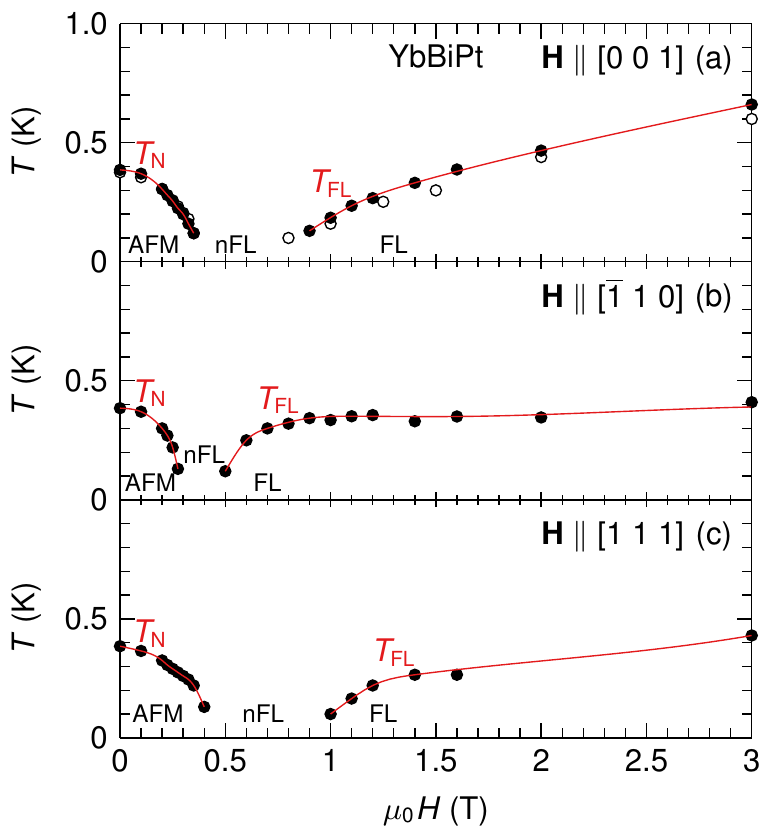}
	\caption{Magnetic phase diagrams based on resistivity data for $\mathbf{I}\perp\mathbf{H}$ with $\mathbf{H}\parallel[0~0~1]$ (a), $[\bar{1}~1~0]$ (b), or $[1~1~1]$ (c).  Open symbols are from Ref.~\onlinecite{Mun_2013} and closed symbols are new data. Lines are guides to the eye.  AFM stands for antiferromagnetic, nFL labels the non-Fermi-liquid region, characterized by $\rho\sim T^{1.5}$, and FL marks the Fermi-liquid region, characterized by $\rho\sim T^{2}$.  $T_{\text{N}}$ labels the N\'{e}el temperature and $T_{\text{FL}}$ marks the boundary of the Fermi-liquid region.}
	\label{Fig2}
\end{figure}

Zero-field resistivity data for five representative samples are plotted in Fig.~\ref{Fig1}(a) along with data from Ref.~\onlinecite{Mun_2013}.  The legend indicates the direction of \textbf{I} for each dataset.  A jump in $\rho(T)$ occurs at $T_{\text{N}}$ in each curve which is little affected by the direction of \textbf{I}.  Figures~\ref{Fig1}(b)--\ref{Fig1}(d) show $\rho(T)$ for specific samples with $\mathbf{H}$ along either $[0~0~1]$, $[\bar{1}~1~0]$, or $[1~1~1]$ and $\mathbf{I}\perp\mathbf{H}$.  The jump at $T_{\text{N}}$ generally shifts to lower $T$ with increasing $H$, however, the amount by which it shifts depends on the field's direction.  The change in $T_{\text{N}}$ with $H$ is determined in the same way as in Ref.~\onlinecite{Mun_2013} and is shown by the AFM boundaries in Fig.~\ref{Fig2} for different directions of $\mathbf{H}$.

Figure~\ref{Fig1}(e) shows the transverse magnetoresistivity (i.e.\ the longitudinal resistivity measured perpendicular to $\mathbf{H}$), $\rho(H)$, at $T=0.1$~K for $\mathbf{H}\parallel[0~0~1]$, $[\bar{1}~1~0]$, and $[1~1~1]$.  Below $\mu_{0}H=0.4$~T, $\rho(H)$ rapidly changes for all three field directions.  At higher $H$, depending on the field direction, $\rho(H)$ shows either a local minimum or levels off near $\mu_{0}H\approx 1$~T.  A broad local minimum occurs at $\mu_{0}H\approx 3$~T for $\mathbf{H}\parallel[0~0~1]$ and a broad local maximum is seen at $\approx 5$~T and $\approx 4$~T for $\mathbf{H}\parallel[\bar{1}~1~0]$ and $\mathbf{H}\parallel[1~1~1]$, respectively. Clear quantum oscillations are present for all three directions at  high field.  The oscillations are analyzed in Ref.~\onlinecite{Mun_2015} for $\mathbf{H}\parallel[0~0~1]$ and $[1~1~1]$.

\begin{figure}
	\centering
	\includegraphics[width=1.0\linewidth]{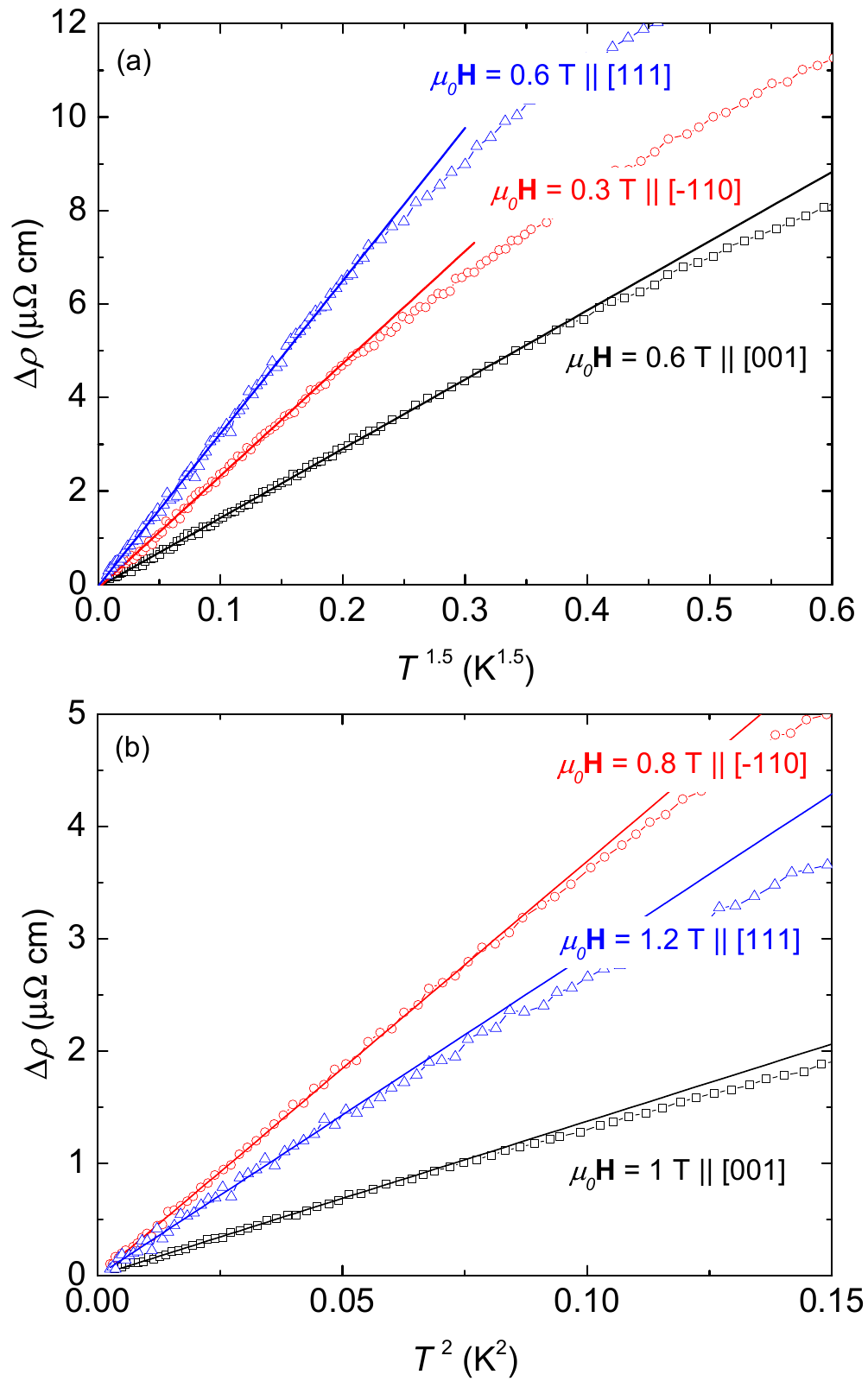}
	\caption{(a) $\Delta\rho(T)=\rho(T)-\rho(T=0$~K$)$ versus $T^{1.5}$ and (b) $\Delta\rho(T)$ versus $T^{2}$ for $\mathbf{H}\parallel[0~0~1]$, $[\bar{1}~1~0]$, or $[1~1~1]$.   \textbf{I} was applied $\perp\mathbf{H}$, along the directions indicated in Fig.~\ref{Fig1}.  Solid lines are guides to the eye.}
	\label{Fig3}
\end{figure}

\begin{figure}
	\centering
	\includegraphics[width=1.0\linewidth]{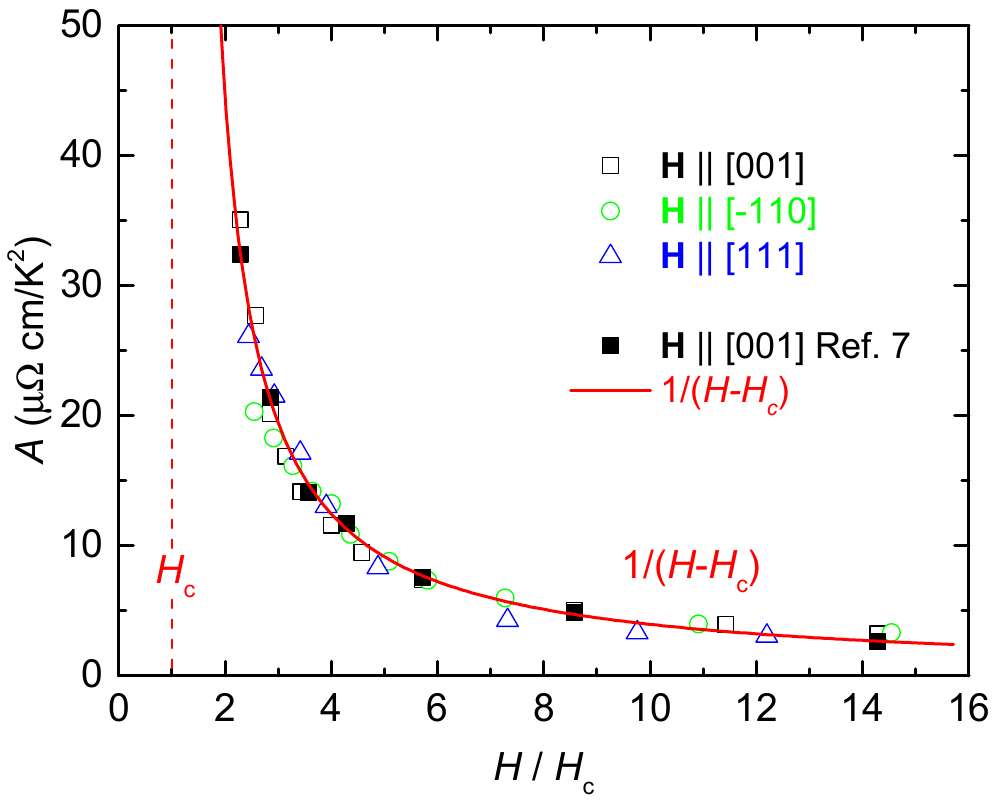}
	\caption{The Fermi-liquid coefficient $A = \Delta\rho(T)/T^{2}$ for $\mathbf{H}\parallel[1~0~0]$,  $[\bar{1}~1~0]$, or $[1~1~1]$.   $\mathbf{I}$ was applied $\perp\mathbf{H}$, along the directions indicated in Fig.~\ref{Fig1}.  Data from Ref.~\onlinecite{Mun_2013} (for $\mathbf{H}\parallel[0~0~1]$) are included for comparison. The solid line is a fit to $A \propto 1/(H - H_{\text{c}})$, where $H_{\text{c}}$ corresponds to the magnetic field at which $T_{\text{N}}$ extrapolates to $0$~K.  Note that the values of $A$ are for values of $H$ and $T$ corresponding to the FL region.}
	\label{Fig4}
\end{figure}

The phase diagrams in Fig.~\ref{Fig2} illustrate that $T_{\text{N}}$ is suppressed to $T=0$~K for some value of field $\mu_{0}H_{c}$, and that a nFL region exists for all three field directions.  By extrapolating the AFM boundary to $T=0$~K, we find that $\mu_{0}H_{\text{c}}=0.37$,  $0.28$, and $0.42$~T for $\mathbf{H}\parallel[0~0~1]$, $[\bar{1}~1~0]$, and $[1~1~1]$, respectively.  Figure~\ref{Fig3} presents fits made to $\Delta\rho(T)\sim T^{1.5}$ [Fig.~\ref{Fig3}(a)] and $\Delta\rho(T)\sim T^{2}$ [Fig.~\ref{Fig3}(b)] for different directions of $\mathbf{H}$, where $\Delta\rho(T)=\rho(T)-\rho(T=0\text{~K})$.  These plots demonstrate how the nFL [$\Delta\rho(T)=AT^{1.5}$] and FL [$\Delta\rho(T)=AT^{2}$] regions in Fig.~\ref{Fig2} are defined via the $\rho(T,H)$ data \cite{Mun_2013}, and the line in Fig.~\ref{Fig2} labeled $T_{\text{FL}}$  marks the upper limit for which $\rho\sim T^{2}$.

Figure~\ref{Fig4} shows the Fermi-liquid coefficient $A$ versus $H/H_{\text{c}}$ for the different directions of $\mathbf{H}$. $A$ was determined from fits to $\Delta\rho(T)=AT^{2}$ within the FL region, and  $A\sim 1/(H-H_{\text{c}})$ for all three field directions.  As discussed previously\cite{Mun_2013}, the tendency of $A$ to diverge as $\mu_{0}H \rightarrow \mu_{0}H_{c}$ provides evidence for the quasiparticle effective mass being enhanced due to quantum fluctuations associated with a QCP.

\subsection{Magnetic neutron diffraction with $\mathbf{H}\bm{\parallel[\bar{1}~1~0]}$} \label{sub3_B}

Much of the neutron diffraction data given in this paper are for $\mathbf{H}\parallel[\bar{1}~1~0]$ rather than the $\mathbf{H}\parallel[0~0~1]$ configuration used for the thermodynamic and transport measurements in Ref.~\onlinecite{Mun_2013}.  This is due to constraints on the neutron diffraction experiments imposed by applying a field along $[0~0~1]$.  In particular, access to multiple magnetic diffraction peaks corresponding to $\bm{\tau}=(\frac{1}{2},\frac{1}{2},\frac{1}{2})$ and $\bm{\mu}\parallel\bm{\tau}$  on the spectrometers used required the sample's $(h,h,l)$ plane to lie horizontal.  This means that a magnet supplying a horizontal field would be necessary to record data for $\mathbf{H}\parallel[0~0~1]$.  We performed some measurements using a horizontal-field magnet and determined that it too greatly limited neutron access for detailed studies of the magnetic diffraction peaks' lineshapes.  This was due to the magnet and its supporting structures blocking or attenuating the neutron beam and the weak and broad nature of the peaks.  On the other hand, performing experiments with $\mathbf{H}\parallel[\bar{1}~1~0]$ allowed for a vertical-field magnet to be used, which provided for much more neutron access to the sample.

\subsubsection{$\mathbf{Q}=(\frac{1}{2},\frac{1}{2},\frac{3}{2})$}
\begin{figure}
	\centering
	\includegraphics[width=1.0\linewidth]{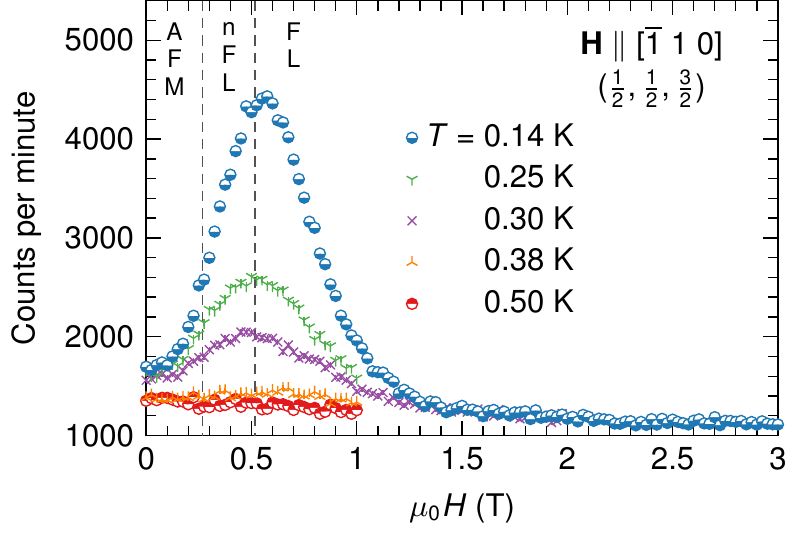}
	\caption{Intensity of the scattering at $(\frac{1}{2},\frac{1}{2},\frac{3}{2})$ versus magnetic field for various temperatures and $\mathbf{H}\parallel[\bar{1}~1~0]$.  Data were taken on SPINS.  Vertical dotted lines mark the boundaries identified in Fig.~\ref{Fig2}(b) at $T=0.14$~K determined from resistivity data.}
	\label{Fig5}
\end{figure}

\begin{figure}
	\centering
	\includegraphics[width=1.0\linewidth]{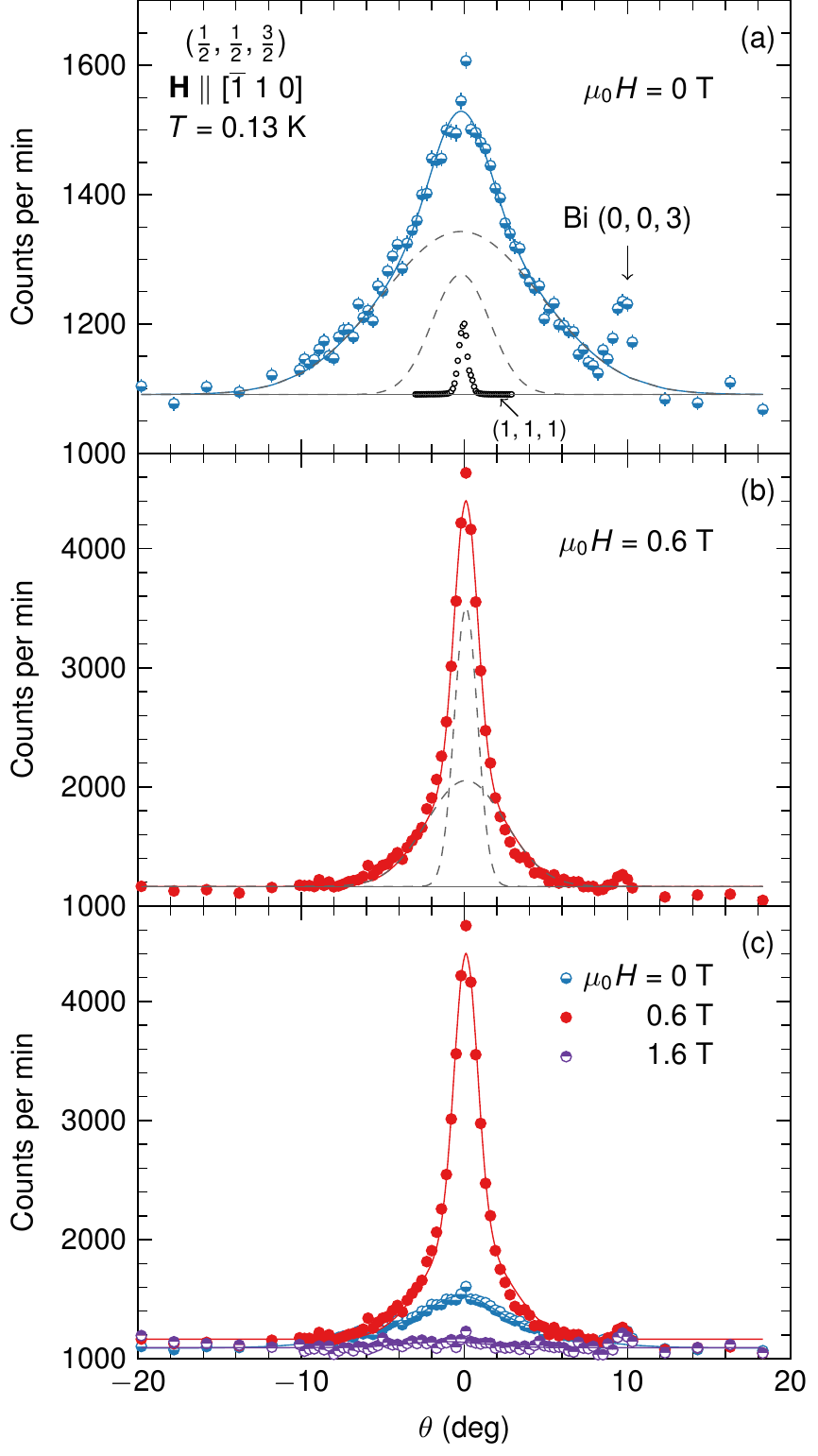}
	\caption{Rocking curves for the $(\frac{1}{2},\frac{1}{2},\frac{3}{2})$ magnetic diffraction peak at $T=0.13$~K for $\mu_{0}H=0$ (a) and $0.6$~T (b) applied parallel to $[\bar{1}~1~0]$.  Data were taken on SPINS using $\lambda=5.504$~\AA\ neutrons.  Solid curves are fits to the two-Gaussian lineshape described in the text, and the lineshape's components are shown by dashed curves.  The small peak at $\approx10$\degree\ is due to residual Bi flux from the crystal synthesis process and was masked while performing fits.  Data for $\mu_{0}H=0$, $0.6$, and $1.6$~T are plotted altogether in (c).  A scaled rocking curve for the $(1,1,1)$ structural Bragg peak at $T=0.13$~K is also shown in (a).}
	\label{Fig6}
\end{figure}

Figure~\ref{Fig5} shows the magnetic field dependence of the scattering intensity of the $(\frac{1}{2},\frac{1}{2},\frac{3}{2})$ magnetic diffraction peak at various temperatures with $\mathbf{H}\parallel[\bar{1}~1~0]$.  Data were recorded on SPINS.  For $T\le0.30$~K, the peak's intensity grows monotonically with increasing field and reaches a maximum at $\mu_{0}H=0.55(5)$~T for $T=0.14$~K, which is just past the FL boundary at $\mu_{0}H=0.52$~T.  There is no clear feature associated with the AFM-nFL boundary identified in Fig.~\ref{Fig2}(b).  The intensity maximum diminishes and shifts to lower $H$ with increasing $T$ and is no longer discernible at $T=0.38$~K. Nevertheless, the scattering intensity is still slightly higher at $T=0.38$~K than at $0.5$~K for $0.3\alt\mu_{0}H\alt1$~T.

Since we previously showed that the narrow-Gaussian component of the $(\frac{1}{2},\frac{1}{2},\frac{3}{2})$ diffraction peak appears at $T_{\text{N}}$ for $\mu_{0}H=0$~T \cite{Ueland_2014}, the presence of the intensity maximum for $T<T_{\text{N}}$ suggests that it is associated with the magnetic diffraction peak's narrow component.   On the other hand, the maximum occurs at a higher field than expected for the AFM boundary given in Fig.~\ref{Fig2}(b). This is shown for $T=0.14$~K by the dashed vertical lines in Fig.~\ref{Fig5}, which mark the boundaries given in Fig.~\ref{Fig2}(b) for $T=0.14$~K and indicate that the maximum occurs near the FL boundary.  To gain more insight into the field dependence of the magnetic diffraction peaks, we next present data that detail the $(\frac{1}{2},\frac{1}{2},\frac{3}{2})$ peak's lineshape.

\begin{figure}
	\centering
	\includegraphics[width=1.0\linewidth]{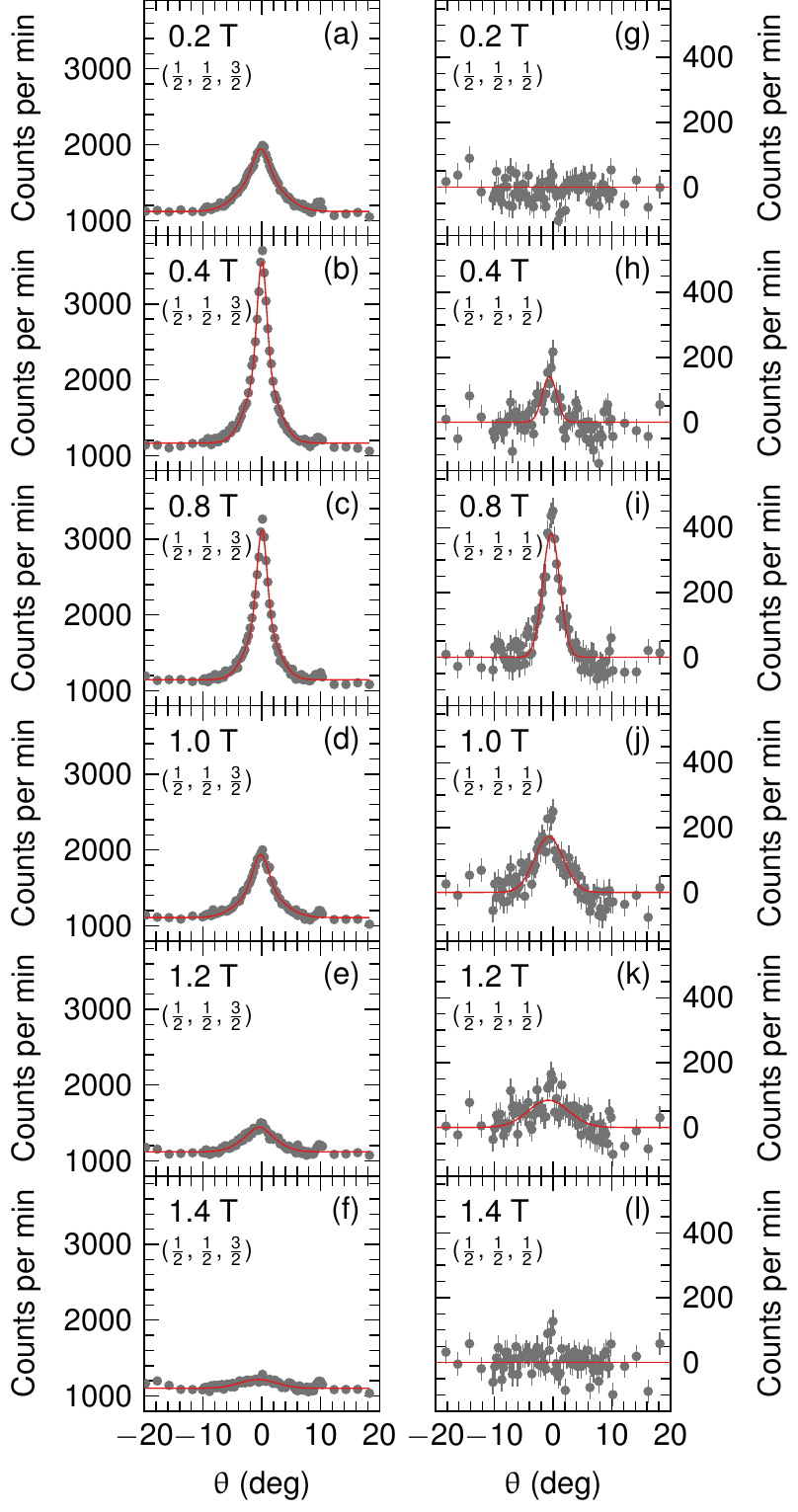}
	\caption{Rocking scan data from SPINS ($\lambda=5.504$~\AA) for the $(\frac{1}{2},\frac{1}{2},\frac{3}{2})$ (a)--(f) and $(\frac{1}{2},\frac{1}{2},\frac{1}{2})$ (g)--(l) reciprocal-lattice positions taken at $T=0.13$ and $0.14$~K, respectively, for various values of $\mathbf{H}\parallel[\bar{1}~1~0]$.  Data in (g)--(l) are subtracted by data for $\mu_{0}H=0$~T.  The solid curves are fits to either a two-Gaussian [$(\frac{1}{2},\frac{1}{2},\frac{3}{2})$] or Gaussian [$(\frac{1}{2},\frac{1}{2},\frac{1}{2})$] lineshape.}
	\label{Fig7}
\end{figure}

\begin{figure}
	\centering
	\includegraphics[width=1.0\linewidth]{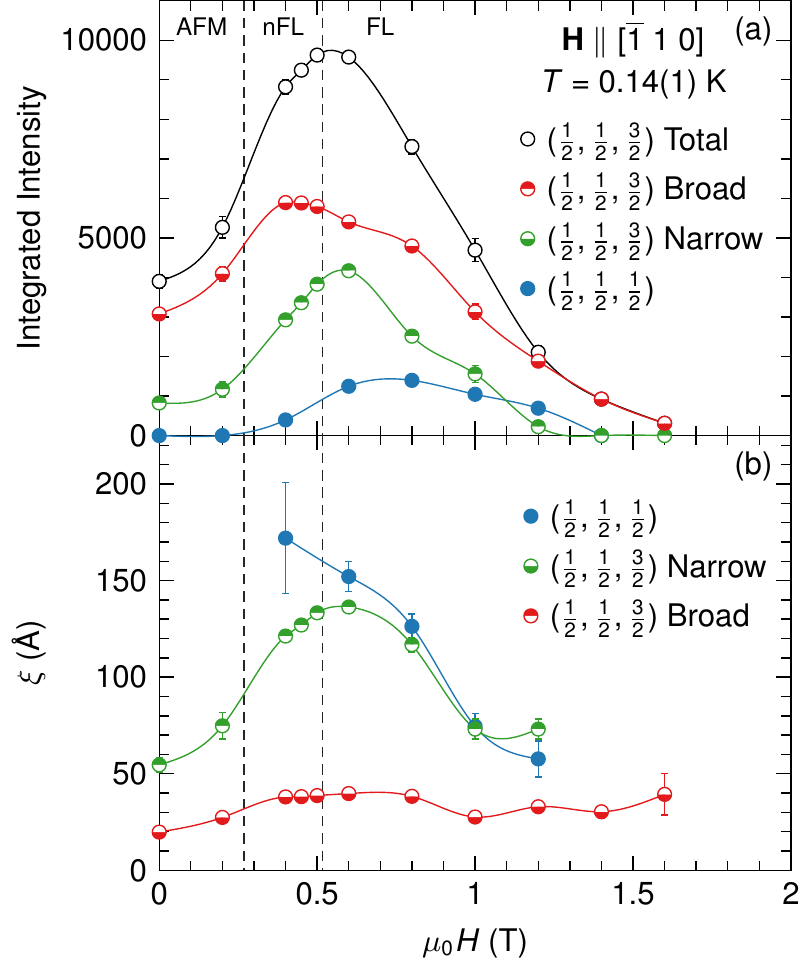}
	\caption{The integrated intensities (a) and magnetic-correlation lengths (b) versus field for $\mathbf{H}\parallel[\bar{1}~1~0]$ and $T=0.14(1)$~K from fits to the rocking curves for the $(\frac{1}{2},\frac{1}{2},\frac{3}{2})$ and $(\frac{1}{2},\frac{1}{2},\frac{1}{2})$ magnetic diffraction peaks shown in Figs.~\ref{Fig6}, \ref{Fig7} and \ref{Fig11}. Narrow and broad refer to the components of the two-Gaussian lineshape fit to the $(\frac{1}{2},\frac{1}{2},\frac{3}{2})$ peak, and total refers to the sum of the integrated intensities of the two components.  Vertical dotted lines mark the boundaries identified in Fig.~\ref{Fig2}(b) at $T=0.14$~K determined from resistivity data.  Solid lines are guides to the eye.}
	\label{Fig8}
\end{figure}

Figure~\ref{Fig6} shows SPINS rocking scan data for the $(\frac{1}{2},\frac{1}{2},\frac{3}{2})$ magnetic diffraction peak at $T=0.13$~K and $\mu_{0}H=0$, $0.6$, and $1.6$~T.  Similar to our previous report \cite{Ueland_2014}, the peak is fit to a two-Gaussian lineshape composed of broad- and narrow-Gaussian components and a constant offset.  The fits in Fig.~\ref{Fig6}(a) for $\mu_{0}H=0$~T give FWHM for the broad and narrow components of $11.5(5)$\degree and $4.2(3)$\degree, respectively, and the fits in Fig.~\ref{Fig6}(b) for $\mu_{0}H=0.6$~T give FWHM of $5.7(1)$\degree and $1.67(3)$\degree, respectively.  Hence, both of the peak's components are sharper for $\mu_{0}H=0.6$~T.  This means that their associated magnetic-correlation lengths are larger than for $\mu_{0}H=0$~T.

On the other hand, Fig.~\ref{Fig6}(a) also shows scaled data for the $(1,1,1)$ structural Bragg peak.  This peak occurs at a value of $2\theta$ (scattering angle) that is only $5$\degree\ higher than that for the $(\frac{1}{2},\frac{1}{2},\frac{3}{2})$  peak, and should give a measure of the experimental resolution which is close to the resolution for the $(\frac{1}{2},\frac{1}{2},\frac{3}{2})$ position.  An estimation of the resolution based on a calculation using the \textsc{dave} software package gives an expected FWHM of $\approx0.6$\degree\ for both peaks.  The FWHM of the $(1,1,1)$ Bragg peak determined from the measurement is $0.70(1)$\degree, which is larger than but close to the estimated resolution.  Since the FWHM of the $(1,1,1)$ structural peak is much smaller than the FWHM of either component of the $(\frac{1}{2},\frac{1}{2},\frac{3}{2})$ magnetic peak for both $\mu_{0}H=0$ and $0.6$~T, we conclude that short-range AFM persists at $0.6$~T.

The $(\frac{1}{2},\frac{1}{2},\frac{3}{2})$ peak is almost suppressed for $\mu_{0}H=1.6$~T, and Fig.~\ref{Fig6}(c) compares scans taken at $T=0.13$~K for $0$, $0.6$, and $1.6$~T.  The value for $\mu$ corresponding to the total integrated intensity of the peak for $\mu_{0}H=0$~T and $T=0.13$~K is $\mu=0.76(6)~\mu_{\text{B}}/$Yb, which is similar to the value of $0.8~\mu_{\text{B}}/$Yb previously reported \cite{Ueland_2014}.  Figures~\ref{Fig7}(a)--\ref{Fig7}(f) exhibit data for other values of $H$.

Figure~\ref{Fig8} details the changes to the $(\frac{1}{2},\frac{1}{2},\frac{3}{2})$ peak's lineshape at $T=0.14(1)$~K induced by $\mathbf{H}\parallel[\bar{1}~1~0]$.  The vertical dashed lines mark the AFM and FL boundaries at $T=0.14$~K given in Fig.~\ref{Fig2}(b), which were determined from resistivity data.  Figure~\ref{Fig8}(a) shows that with increasing $H$ the integrated intensities of the peak's narrow and broad components both rise between $\mu_{0}H=0$ and $0.4$~T.  For $\mu_{0}H>0.4$~T, the integrated intensity of the narrow component continues to increase until reaching a maximum near the FL boundary, and then falls to $0$ past $1.2$~T.  The integrated intensity of the broad component gently decreases over $0.4\le\mu_{0}H\alt0.8~T$, and falls more rapidly for $\mu_{0}H>0.8$~T, approaching $0$ at $1.6$~T.

Next, Fig.~\ref{Fig8}(b) shows the magnetic-correlation lengths associated with the peak's broad ($\xi_{\text{B}}$) and narrow ($\xi_{\text{N}}$) components as a function of field.  These lengths were determined after calculating the trajectory of the rocking scan in $Q$ space in terms of reciprocal-lattice units.  The corresponding FWHM of the peak in reciprocal-lattice units was then converted to \AA.   $\xi_{\text{B}}$ increases from $19.8(9)$ to $39.7(9)$~\AA\ between $\mu_{0}H=0$ and $0.6$~T, and fluctuates around a slightly lower value for $0.6\alt \mu_{0}H \le 1.6$~T.   $\xi_{\text{N}}$ grows from $54(4)$ to$136(2)$~\AA\ between $\mu_{0}H=0$ and $0.6$~T, reaching a maximum at $0.6$~T, just past the FL boundary. It decreases between $\mu_{0}H=0.6$ and $1.0$~T, and $\xi_{\text{N}}=73(5)$~\AA\ for both $1.0$ and $1.2$~T.

\begin{figure}
	\centering
	\includegraphics[width=1.0\linewidth]{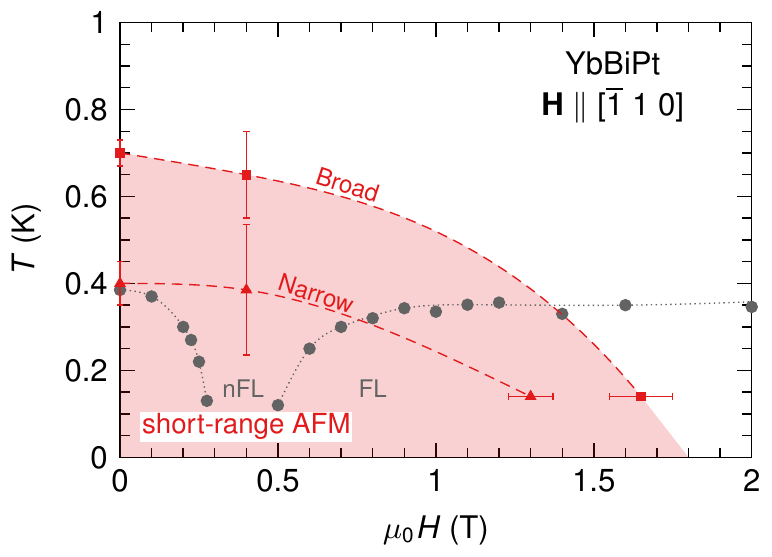}
	\caption{Diagram showing the region for which magnetic diffraction peaks are found for $\mathbf{H}\parallel[\bar{1}~1~0]$.  The AFM and FL boundaries determined via resistivity are also shown.  Squares (triangles) mark the onset temperatures and magnetic fields of the broad (narrow) component of the $(\frac{1}{2},\frac{1}{2},\frac{3}{2})$ magnetic diffraction peak, and the shaded region indicates where short-range AFM exists.  Lines are guides to the eye.  nFL and FL label the non-Fermi-liquid and Fermi-liquid regions, respectively.  }
	\label{Fig9}
\end{figure}

\begin{figure}
	\centering
	\includegraphics[width=1.0\linewidth]{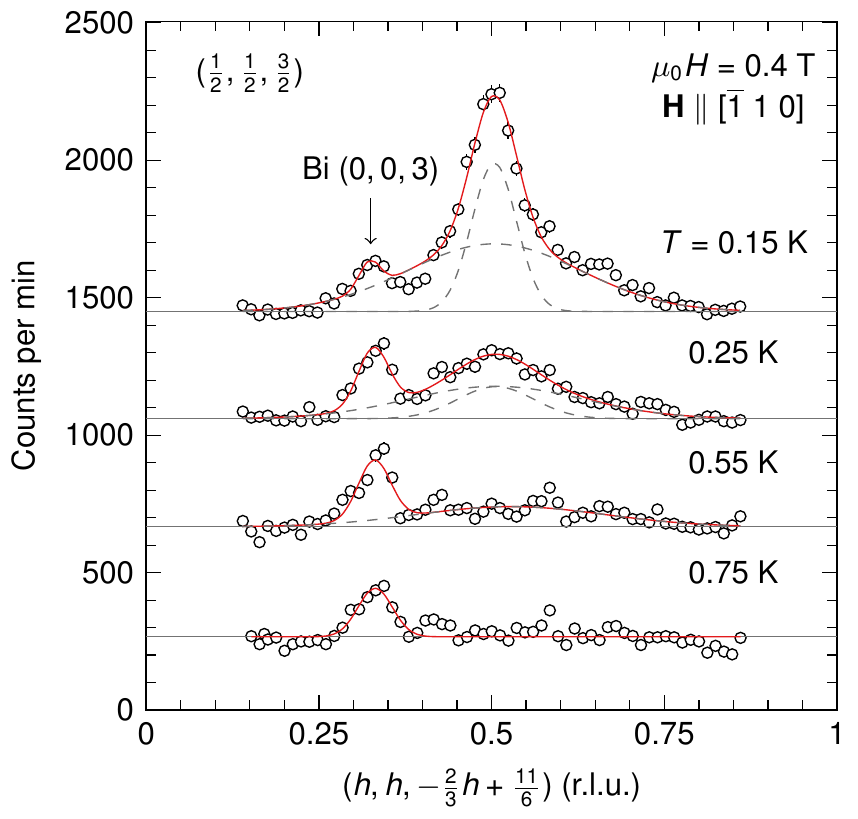}
	\caption{Data from transverse scans taken on FLEXX, using $\lambda=5.464$~\AA\ neutrons, showing the temperature dependence of the $(\frac{1}{2},\frac{1}{2},\frac{3}{2})$ magnetic diffraction peak for $\mu_{0}H=0.4$~T. Data are successively offset by 400 counts per minute.  Solid red curves are fits to the two-Gaussian lineshape described in the text. The lineshape's components are shown by dashed curves.  Bragg peaks due to residual Bi flux from the crystal synthesis process are also marked.}
	\label{Fig10}
\end{figure}

The field dependencies shown in Fig.~\ref{Fig8} are not those we anticipated from Fig.~\ref{Fig2}(b).  Specifically, we expected the integrated intensity and correlation length associated with the $(\frac{1}{2},\frac{1}{2},\frac{3}{2})$ peak to decrease as the AFM-nFL boundary is approached and crossed with increasing field.  Another way of seeing this disagreement is to overlay in Fig.~\ref{Fig9} the region for which the $(\frac{1}{2},\frac{1}{2},\frac{3}{2})$ magnetic diffraction peak exists on top of the phase boundaries for $\mathbf{H}\parallel[\bar{1}~1~0]$ determined via resistivity data and shown in Fig.~\ref{Fig2}(b).  The points at $\mu_{0}H=0.4$~T are from Fig.~\ref{Fig10} (described below), and points from the neutron diffraction study in Ref.~\onlinecite{Ueland_2014} are also incorporated.  The region where the peak exists is shaded, and dashed lines are estimates for the boundaries of its broad and narrow components.    The figure clearly demonstrates disagreement between the neutron diffraction data and the AFM boundary determined by resistivity.  The reason for this disagreement is currently unknown.

To end this subsection, we show in Fig.~\ref{Fig10} data from transverse scans taken across the $(\frac{1}{2},\frac{1}{2},\frac{3}{2})$ diffraction peak for $\mu_{0}H=0.4$~T and various temperatures.  Measurements were made using FLEXX and on a different sample than the one used on SPINS.  Similar to data in Fig.~\ref{Fig6}, the magnetic diffraction peak has narrow and broad Gaussian components for $T=0.15$ and $0.25$~K, however, only the broad component exists at $0.55$~K.  The diffraction peak appears to be completely suppressed at $T=0.75$~K.  This is similar to the temperature dependence of the peak for $\mu_{0}H=0$~T \cite{Ueland_2014}.  In Sec.~\ref{Sec4} we argue that the broadness of the $(\frac{1}{2},\frac{1}{2},\frac{3}{2})$ diffraction peak likely reflects the presence of magnetic domains of short-range AFM order, and that within the AFM phase determined from resistivity data, a changing magnetic field changes the domains' populations. Once a high enough field is reached, $\bm{\mu}$ then reorients towards $\mathbf{H}$.

\subsubsection{$\mathbf{Q}=(\frac{1}{2},\frac{1}{2},\frac{1}{2})$}
\begin{figure}
	\centering
	\includegraphics[width=1.0\linewidth]{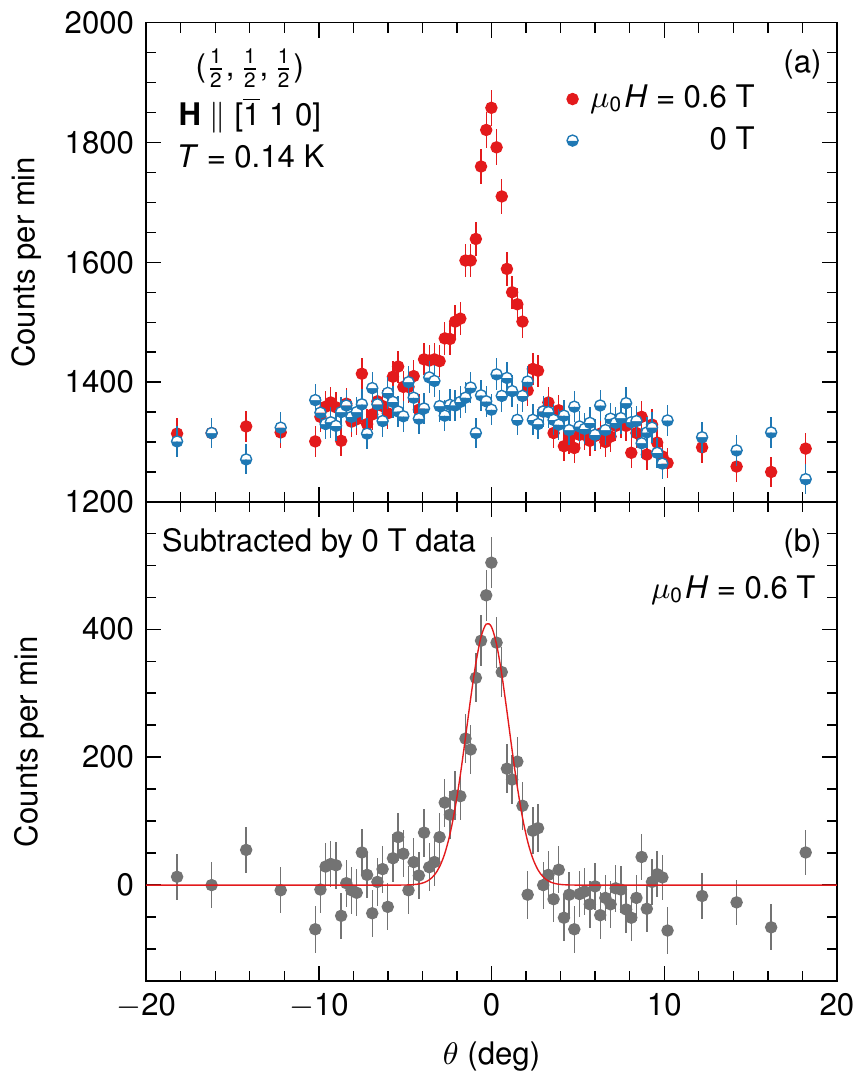}
	\caption{(a) Rocking curves taken on SPINS, using $\lambda=5.504$~\AA\ neutrons, for the $(\frac{1}{2},\frac{1}{2},\frac{1}{2})$ reciprocal-lattice position at $T=0.14$~K for $\mu_{0}H=0$ and $0.6$~T applied $\parallel[\bar{1}~1~0]$. (b) Difference of the $\mu_{0}H=0.6$ and $0$~T data.  The solid line is a fit to a Gaussian lineshape.}
	\label{Fig11}
\end{figure}

Rocking curves for the $(\frac{1}{2},\frac{1}{2},\frac{1}{2})$ reciprocal-lattice position at $T=0.14$~K for $\mu_{0}H=0$ and $0.6$~T applied along $[\bar{1}~1~0]$ are shown in Fig.~\ref{Fig11}(a).  These data were taken on SPINS.  No peak is observed for $\mu_{0}H=0$~T, as expected, since $\bm{\mu}\parallel\bm{\tau}$ and neutron scattering is sensitive only to the component of $\bm{\mu}\perp\mathbf{Q}$ \cite{Lynn_1994}. However, a peak is found for $\mu_{0}H=0.6$~T.  This means that $\bm{\mu}$ is rotated away from $\bm{\tau}$ for this value of field.  To account for the $\theta$ dependence of the background, which is likely dominated by absorption due to the sample holder, we consider the $\mu_{0}H=0$~T data to arise from a nonmagnetic background and subtract them from the $\mu_{0}H=0.6$~T data.  The result is shown in Fig.~\ref{Fig11}(b), wherein the solid line is a fit to a Gaussian lineshape with a FWHM of $2.9(1)$\degree.  This is much larger than the estimated experimental resolution of $1.1$\degree\ found from a calculation using the \textsc{dave} software package. The FWHM of the peak corresponds to a magnetic-correlation length of $\xi_{\frac{1}{2}}=152(8)$~\AA, which is slightly larger than $\xi_{\text{N}}(\mu_{0}H=0.6~\text{T})$.  

Rocking scan data for other values of field are given in Figs.~\ref{Fig7}(g)--\ref{Fig7}(l), and Fig.~\ref{Fig8} plots the field dependencies of the peak's fit parameters.  Their magnetic-field dependencies qualitatively follow those of the narrow component of the $(\frac{1}{2},\frac{1}{2},\frac{3}{2})$ magnetic diffraction peak for $\mu_{0}H\ge0.4$~T.   Note that even though both the narrow-component of the $(\frac{1}{2},\frac{1}{2},\frac{3}{2})$ peak and the $(\frac{1}{2},\frac{1}{2},\frac{1}{2})$  peak have FWHM which are larger than the resolution, the FWHM are only a couple of degrees of rocking angle.  Thus, we expect that the FWHM approximately correspond to those that would be obtained from transverse scans.

\subsection{Magnetic neutron diffraction with $\mathbf{H}\bm{\parallel[0~0~1]}]$ or $\bm{[1~1~0]}$}

\begin{figure}
	\centering
	\includegraphics[width=1.0\linewidth]{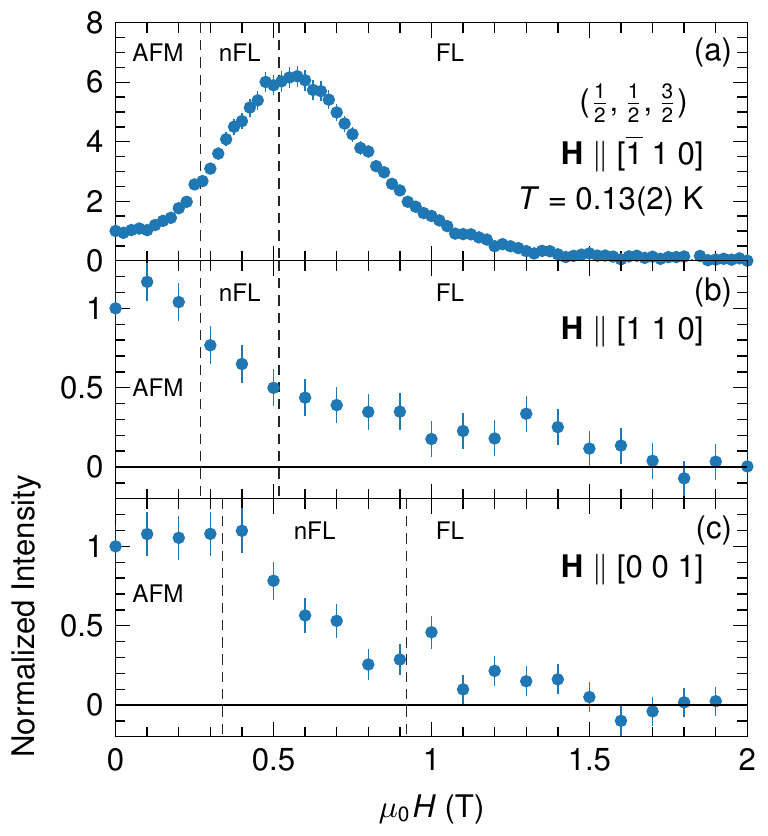}
	\caption{Intensity of the $(\frac{1}{2},\frac{1}{2},\frac{3}{2})$ magnetic diffraction peak versus magnetic field at $T=0.13(2)$~K for a vertical field applied along $[\bar{1}~1~0]$ (a), or a horizontal field applied within the scattering plane along either $[1~1~0]$ (b) or $[0~0~1]$ (c).  Data are scaled to $1$ at $\mu_{0}H=0$~T and $0$ at $2$~T, as described in the text.  Data in (a) are the SPINS data shown in Fig.~\ref{Fig5}, and data in (b) and (c) were taken on E-$4$.  Vertical dotted lines correspond to the boundaries at $T=0.14$~K given in Figs.~\ref{Fig2}(b) and \ref{Fig2}(a).}
	\label{Fig12}
\end{figure}

The effect of the direction of $\mathbf{H}$ on the $(\frac{1}{2},\frac{1}{2},\frac{3}{2})$ magnetic diffraction peak was investigated by recording data on E-$4$ while applying $\mathbf{H}$ within the $(h,h,l)$ scattering plane along either $[0~0~1]$ or $[1~1~0]$.  Figure~\ref{Fig12} shows the peak's intensity versus $H$ at $T=0.13(2)$~K, along with the data from SPINS for $\mathbf{H}\parallel[\bar{1}~1~0]$ (i.e.\ applied perpendicular to the scattering plane) originally shown in Fig.~\ref{Fig5}.   For easier comparison, the datasets are plotted according to the equation
\begin{equation}
\frac{S(\mu_{0}H)-S(\mu_{0}H=2~\text{T})}{S(\mu_{0}H=0~\text{T})-S(\mu_{0}H=2~\text{T})},
\end{equation}
where $S$ stands for the scattering intensity, and scattering recorded for $\mu_{0}H\ge2$~T is assumed to be due to a constant nonmagnetic background.  This scaling is consistent with Fig.~\ref{Fig8}(a), because Fig~\ref{Fig8}(a) shows the scattering intensity approaching $0$ at $\mu_{0}H\approx1.6$~T for $\mathbf{H}\parallel[\bar{1}~1~0]$ at $T=0.14$~K.

In contrast to the $\mathbf{H}\parallel[\bar{1}~1~0]$ data, Figs.~\ref{Fig12}(b) and \ref{Fig12}(c) show that the magnetic scattering intensity generally decreases throughout the nFL and FL regions for both $\mathbf{H}\parallel[1~1~0]$ and $[0~0~1]$.  The reason behind the scattering intensity's dependence on magnetic field direction is discussed below.  For now, we note that data in Fig.~\ref{Fig12}(c) do show a change in behavior near the AFM-nFL boundary determined in Ref.~\onlinecite{Mun_2013}, however, the magnetic scattering persists well past this boundary.

\section{Discussion}\label{Sec4}

\subsection{Antiferromagnetic domains and the reciprocal lattice}\label{sub4_A}
\begin{figure}
	\centering
	\includegraphics[width=1.0\linewidth]{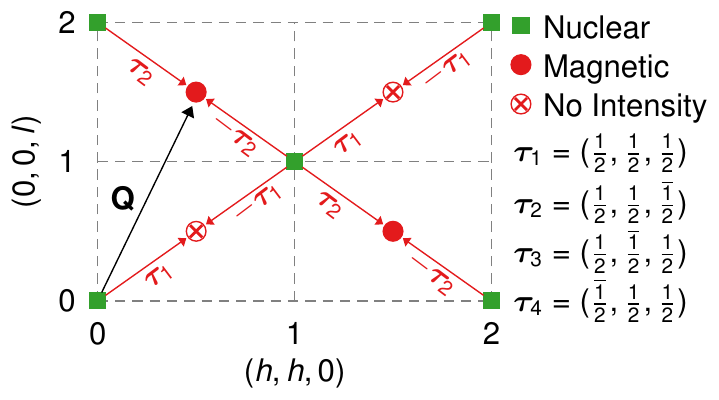}
	\caption{Diagram of the $(h,h,l)$ reciprocal-lattice plane showing structural (squares) and magnetic (circles) Brillouin zone centers, as well as the AFM propagation vectors connecting them.  $\mathbf{Q}=(\frac{1}{2},\frac{1}{2},\frac{3}{2})$ is also shown.  A circle with an $\times$ in its interior indicates a magnetic zone center with zero diffraction intensity due to the associated magnetic moments being oriented along $\bm{\tau}$, as occurs for $\mu_{0}H=0$~T.}
	\label{Fig13}
\end{figure}

We begin our discussion by examining the diagram of YbBiPt's $(h,h,l)$ reciprocal-lattice plane given in Fig.~\ref{Fig13}, which shows the connection between the structural and magnetic reciprocal lattices for $\bm{\tau}=(\frac{1}{2},\frac{1}{2},\frac{1}{2})$. We assume collinear AFM order with $\bm{\mu}\parallel\bm{\tau}$, as described in Ref.~\onlinecite{Ueland_2014}. The positions of the magnetic diffraction peaks and the body-centered-cubic reciprocal-lattice symmetry leads to $4$ equivalent AFM propagation vectors: $\bm{\tau}_{1}=(\frac{1}{2},\frac{1}{2},\frac{1}{2})$, $\bm{\tau}_{2}=(\frac{1}{2},\frac{1}{2},\frac{\bar{1}}{2})$, $\bm{\tau}_{3}=(\frac{1}{2},\frac{\bar{1}}{2},\frac{1}{2})$, and $\bm{\tau}_{4}=(\frac{\bar{1}}{2},\frac{1}{2},\frac{1}{2})$.  Each vector may be thought of as representing AFM domains with an ordered moment $\bm{\mu_{i}}$ oriented along $\bm{\tau}_{i}$.  For example, Fig.~\ref{Fig13} shows that the $(\frac{1}{2},\frac{1}{2},\frac{3}{2})$ and $(\frac{1}{2},\frac{1}{2},\frac{1}{2})$ positions are connected to the  structural reciprocal lattice by $\bm{\tau}_{2}$ and $\bm{\tau}_{1}$, respectively.  Thus, any magnetic Bragg peaks found via neutron diffraction at these positions would be associated with either $\bm{\tau}_{2}$ or $\bm{\tau}_{1}$ magnetic domains. In the case of ideal long-range AFM order, neutron diffraction experiments would find similar resolution limited magnetic Bragg peaks at symmetry equivalent reciprocal-lattice positions corresponding to each domain.  For the case of YbBiPt, we find broad magnetic diffraction peaks at these positions which indicate short-range AFM order.  The magnetic-correlation length determined by fitting such a diffraction peak may be interpreted as being the average size of the $\bm{\tau}_{i}$ domains being probed.

We next consider that the expected response of a collinear AFM to an increasing magnetic field includes changes to the populations (hence the average sizes) of magnetic domains and the eventual reorientation of $\bm{\mu}$ towards $\mathbf{H}$.  For low values of $H$, magnetic domains with the largest component of $\bm{\mu}$ perpendicular to $\mathbf{H}$ are energetically favorable, and the sizes of these domains will increase with increasing $H$ at the expense of unfavorable domains.  This will occur until a sufficiently high $H$ is reached for which it becomes energetically favorable for $\bm{\mu}$ to reorient towards $\mathbf{H}$.  Such a response has been detailed, for example, through neutron diffraction results for Cr \cite{Fawcett_1988} and UPdSn \cite{Werner_1967,Nakotte_1993, Nakotte_1998}.  For YbBiPt, we will show in Sec.~\ref{sub4_B} that at $T=0.13(2)$~K application of a field within the AFM region changes the relative populations of magnetic domains and that the field-induced domain repopulation continues into the nFL region.  In Sec.~\ref{sub4_C} we will show that $\bm{\mu}$ starts to reorient with increasing $H$ approximately halfway through the nFL region, and turns away from $\bm{\tau}$ for $0.4\alt \mu_{0}H\approx1.2$~T.

\subsection{Changes to the populations of magnetic domains}\label{sub4_B}

For $\mathbf{H}\parallel[\bar{1}~1~0]$ (i.e.\ $\perp\bm{\tau}_{1}$ and $\bm{\tau}_{2}$), magnetic domains with $\bm{\mu}\parallel\bm{\tau}_{1}$ or $\bm{\tau}_{2}$ are energetically favorable and will grow in size with increasing $H$ as domains corresponding to $\bm{\tau}_{3}$ and $\bm{\tau}_{4}$ shrink. The field dependence of the $(\frac{1}{2},\frac{1}{2},\frac{3}{2})$ magnetic diffraction peak at $T=0.14$~K appears to follow this behavior throughout the AFM and nFL regions:  Figure~\ref{Fig8}(a) shows that the integrated intensities of both components of the peak increase between $\mu_{0}H=0$ and $0.55$~T, and Fig.~\ref{Fig8}(b) shows that both $\xi_{\text{N}}$ and $\xi_{\text{B}}$  also increase over this field range.  Thus, the average size of the $\bm{\tau}_{2}$ domains grows, and either the number of ordered moments within them increases or $\bm{\mu}_{2}$ becomes larger.  Note that Figs.~\ref{Fig5} and \ref{Fig12}(a) show that the $H$ dependence of the peak's intensity reflects the field-induced changes to its FWHM and integrated intensity.  Assuming that this holds for other directions of $\mathbf{H}$, we next consider data for $\mathbf{H}\parallel[1~1~0]$ and $[0~0~1]$ using the intensity versus field data in Figs.~\ref{Fig12}(b) and \ref{Fig12}(c).

For $\mathbf{H}\parallel[1~1~0]$ (i.e.\ $\perp\bm{\tau}_{3}$ and $\bm{\tau}_{4}$), domains corresponding to $\bm{\tau}_{3}$ and $\bm{\tau}_{4}$ are energetically favorable, and Fig.~\ref{Fig12}(b) shows that the diffraction peak's intensity begins to decrease near the same value of field at which the data in Fig.~\ref{Fig12}(a) begin to increase.  This is expected since when $H$ increases the AFM domain being probed shrinks for $\mathbf{H}\parallel[1~1~0]$ and grows for $\mathbf{H}\parallel[\bar{1}~1~0]$.  A magnetic field applied parallel to $[0~0~1]$ makes the same angle with all four $\bm{\tau}_{i}$s, which means that none of the magnetic domains are more energetically favorable than the others.  Thus Fig.~\ref{Fig12}(c) shows that the scattering intensity at $(\frac{1}{2},\frac{1}{2},\frac{3}{2})$ is constant for $\mu_{0}H\alt0.4$~T.  This encompasses most of the region for which the intensity in Fig~\ref{Fig12}(a) grows, and, as we discuss in the next subsection, is the region for which no reorientation of $\bm{\mu}$ occurs.

It is clear that the intensity of the $(\frac{1}{2},\frac{1}{2},\frac{3}{2})$ magnetic diffraction peak does not disappear upon crossing the AFM boundary for any of the three field directions discussed above, and Fig.~\ref{Fig8}(b) also shows that its magnetic-correlation length behaves differently than typically expected for an AFM phase transition.  In particular, a second-order paramagnetic to AFM transition would involve the growth of dynamic AFM correlations on the paramagnetic side of the transition as the phase  boundary is approached.  These correlations would evolve into magnetic diffraction peaks at reciprocal-lattice positions corresponding to $\bm{\tau}$ upon crossing into the AFM phase.  For example, data for CeCu$_{5.8}$Au$_{0.2}$ taken for an increasing magnetic field at $T=0.06$~K~($<T_{\text{N}}$) show broadening of its magnetic Bragg peaks due to a transition out of its AFM ordered state at $\mu_{0}H_{c}=0.35$~T \cite{Stockert_2007,Lohneysen_2007}.  Such broadening would correspond to a decrease in the magnetic-correlation length as the material loses its AFM order.

In the case of YbBiPt, instead of $\xi_{\text{B}}(H)$ and $\xi_{\text{N}}(H)$ being maximum in the AFM phase, we find that both \emph{increase} as the AFM boundary is approached with increasing field at $T=0.14$~K, and that they are both largest within the nFL phase.  No rapid increases or decreases in $\xi_{\text{B}}(H)$ and $\xi_{\text{N}}(H)$ are observed at either the AFM or FL boundaries.  Thus, we do not observe the typical critical behavior expected for a second-order magnetic transition at either the AFM or FL boundaries.  The absence of such critical behavior at the AFM boundary is particularly surprising because in addition to a jump in $\rho(T)$ at $T_{\text{N}}$ \cite{Mun_2013}, heat capacity, thermal expansion, and magnetostriction data show features at $T_{\text{N}}$ which may be associated with an AFM transition \cite{Mun_2013,Fisk_1991}.  Whereas transport measurements may be affected by scattering associated with magnetic domain boundaries, the existence of domain boundaries is, in general, not expected to greatly affect these three measurements.

\subsection{Reorientation of $\bm{\mu}$ towards $\mathbf{H}$}\label{sub4_C}
Figure~\ref{Fig8} shows that for $\mathbf{H}\parallel[\bar{1}~1~0]$ and $T=0.14$~K a magnetic diffraction peak appears at $(\frac{1}{2},\frac{1}{2},\frac{1}{2})$ within the nFL region and persists into the FL region.  As noted in Sec.~\ref{sub4_A}, reorientation of the moments towards the field direction is expected to occur once a certain threshold value of field is reached.  Here, we assume that the appearance of the $(\frac{1}{2},\frac{1}{2},\frac{1}{2})$ peak signals that $H$ is strong enough to reorient $\bm{\mu}_{1}$ away from $\bm{\tau}_{1}$ and towards $\mathbf{H}$.  Since Fig.~\ref{Fig8} shows that the field-induced changes to the peak's lineshape mimic those that occur for the narrow component of the $(\frac{1}{2},\frac{1}{2},\frac{3}{2})$ peak, which is related to $\bm{\tau}_{2}$, we propose that the angle by which $\bm{\mu}$ rotates out of the scattering plane toward $\mathbf{H}$ may be found by comparing the integrated intensity of the narrow component to the integrated intensity of the $(\frac{1}{2},\frac{1}{2},\frac{1}{2})$ peak.  This is because, for $\mathbf{H}\parallel[\bar{1}~1~0]$, $\mathbf{H}$ should influence the $\bm{\tau}_{1}$ and $\bm{\tau}_{2}$ domains in the same manner, and the ratio of the integrated intensities of magnetic-diffraction peaks corresponding to $\bm{\tau}_{1}$ and $\bm{\tau}_{2}$ is expected to depend only on the underlying magnetic order and moment orientation.  We present this analysis below.

The equation relating the integrated intensity of a magnetic Bragg peak to its structure factor is\cite{Lynn_1994}
\begin{equation}
I=CLp^{2}\mu^{2}\left|F\right|^{2}\langle 1-(\bm{\hat{\mu}}\cdot\bm{\hat{\mathbf{Q}}})^{2} \rangle,\label{Eq1}
\end{equation}
where $\langle~\rangle$ indicates averaging over magnetic domains\cite{Shirane_2002}, $I$ is the integrated intensity of the magnetic diffraction peak, $p=0.2695\times10^{-14}$~m, $L$ is the Lorentz factor \cite{McIntyre_1988,Axe_1983},  $F$ is the magnetic structure factor with its dependence on the angle between $\bm{\mu}$ and $\mathbf{Q}$ factored out, and $C$ is a constant related to the instrument. The Yb$^{3+}$ magnetic form factor is included in $F$.  Equation~\ref{Eq1} allows us to write the ratio of the integrated intensity of the narrow component of the $(\frac{1}{2},\frac{1}{2},\frac{3}{2})$ peak to the integrated intensity of the $(\frac{1}{2},\frac{1}{2},\frac{1}{2})$ peak:
\begin{equation}
\frac{I_{(\frac{1}{2},\frac{1}{2},\frac{3}{2})}}{I_{(\frac{1}{2},\frac{1}{2},\frac{1}{2})}}=\frac{F^{2}_{(\frac{1}{2},\frac{1}{2},\frac{3}{2})}L_{(\frac{1}{2},\frac{1}{2},\frac{3}{2})}\langle 1-(\bm{\hat{\mu}}_{(\frac{1}{2},\frac{1}{2},\frac{3}{2})}\cdot{\hat{\mathbf{Q}}_{(\frac{1}{2},\frac{1}{2},\frac{3}{2})}})^{2}\rangle}{F^{2}_{(\frac{1}{2},\frac{1}{2},\frac{1}{2})}L_{(\frac{1}{2},\frac{1}{2},\frac{1}{2})}\langle 1-(\bm{\hat{\mu}}_{(\frac{1}{2},\frac{1}{2},\frac{1}{2})}\cdot{\hat{\mathbf{Q}}_{(\frac{1}{2},\frac{1}{2},\frac{1}{2})}})^{2}\rangle},\label{Eq2}
\end{equation}
The subscripts in Eq.~\ref{Eq2} refer to the  peaks' positions. 

\begin{figure}
	\centering
	\includegraphics[width=1.0\linewidth]{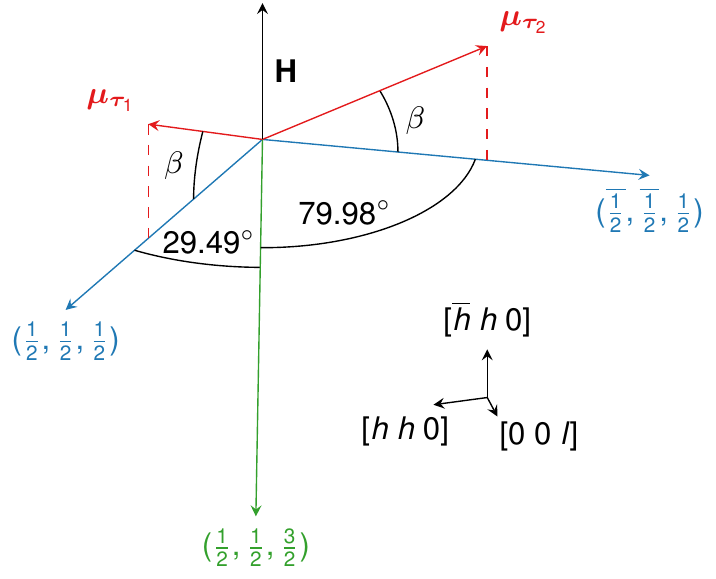}
	\caption{Diagram showing the angles between $\mathbf{H}$, $\mathbf{Q}$, and $\bm{\mu}$ for $\mathbf{Q}=(\frac{1}{2},\frac{1}{2},\frac{3}{2})$ and $(\frac{1}{2},\frac{1}{2},\frac{1}{2})$, and $\mathbf{H}\parallel[\bar{1}~1~0]$.  $\bm{\mu_{\tau_{\text{1}}}}$ and $\bm{\mu}_{\bm{\tau}_{2}}$ refer to moments associated with domains corresponding to $\bm{\tau}_{1}$ and $\bm{\tau}_{2}$, respectively. $\beta$ is the angle by which $\bm{\mu}$ is rotated out of the $(h,~h,~l)$ plane by $\mathbf{H}$.}
	\label{Fig14}
\end{figure}

Next, we consider the specific case of $\mathbf{H}\parallel[\bar{1}~1~0]$, for which magnetic domains corresponding to $\bm{\tau}_{1}$ and $\bm{\tau}_{2}$ are energetically favorable, and assume that $\mathbf{H}$ causes $\bm{\mu}$ to rotate out of the $(h,h,l)$ scattering plane by an angle $\beta$ but the AFM order remains collinear. With the aid of Fig.~\ref{Fig14}, and the facts that the $(\frac{1}{2},\frac{1}{2},\frac{3}{2})$ peak is associated with $\bm{\tau}_{2}$ and the $(\frac{1}{2},\frac{1}{2},\frac{1}{2})$ peak is associated with $\bm{\tau}_{1}$, Eq.~\ref{Eq2} gives
\begin{equation}
\beta=\sin^{-1}\sqrt{\frac{32}{33N-1}},\label{Eq3}
\end{equation}
where
\begin{equation}
N=\frac{I_{(\frac{1}{2},\frac{1}{2},\frac{3}{2})}L_{(\frac{1}{2},\frac{1}{2},\frac{1}{2})}F^{2}_{(\frac{1}{2},\frac{1}{2},\frac{1}{2})}}{I_{(\frac{1}{2},\frac{1}{2},\frac{1}{2})}L_{(\frac{1}{2,}\frac{1}{2},\frac{3}{2})}F^{2}_{(\frac{1}{2},\frac{1}{2},\frac{3}{2})}}.\label{Eq4}
\end{equation}
We assume that the $\bm{\tau}_{1}$ and $\bm{\tau}_{2}$ magnetic domains are equally populated, and remain so with increasing field.

\begin{figure}
	\centering
	\includegraphics[width=1.0\linewidth]{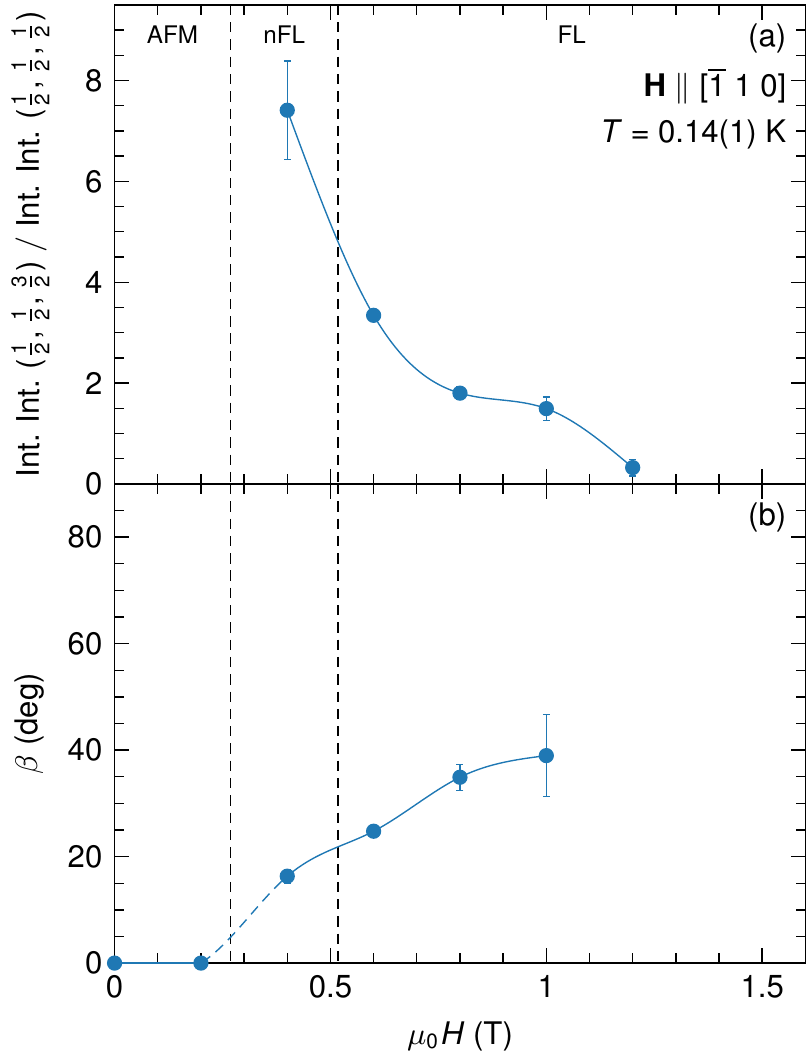}
	\caption{(a) The ratio of the integrated intensity of the narrow component of the $(\frac{1}{2},\frac{1}{2},\frac{3}{2})$ magnetic diffraction peak to the integrated intensity of the $(\frac{1}{2},\frac{1}{2},\frac{1}{2})$ magnetic diffraction peak for $\mathbf{H}\parallel[\bar{1}~1~0]$ and $T=0.14(1)$~K. (b) The angle $\beta$ by which the field causes $\bm{\mu}$ to rotate out of the scattering plane.  A point at $\mu_{0}H=1.2$~T is absent, because $\sqrt{32/(33N-1)}$ is outside of the valid range for $\sin^{-1}$ (see Eqs.~\ref{Eq3} and \ref{Eq4}).  Vertical dotted lines correspond to the boundaries identified in Figs.~\ref{Fig2}(a) and \ref{Fig2}(b) for $T=0.14$~K which were determined from resistivity data. Solid lines are guides to the eye.}
	\label{Fig15}
\end{figure}
 
Figure~\ref{Fig15}(a) shows the ratio of the integrated intensities of the two peaks versus field at $T=0.14$~K and Fig.~\ref{Fig15}(b) shows that $\beta(H)$ at $T=0.14$~K smoothly changes between $\mu_{0}H=0.4$ and $1$~T.  $\bm{\mu}$ points $\approx39$\degree\,out of the $(h,h,l)$ plane by $\mu_{0}H=1$~T, and $\beta(H)$ appears to level off between $\mu_{0}H=0.8$ and $1$~T.  However, it is unclear if the short-range AFM order associated with the narrow component of the  $(\frac{1}{2},\frac{1}{2},\frac{3}{2})$ peak disappears within the FL region before $\bm{\mu}$ reorients parallel to $\mathbf{H}$, because $\beta$ cannot be determined for $\mu_{0}H=1.2$~T due to $(\frac{32}{33N-1})^\frac{1}{2}$ being $>1$.  This is an invalid argument for $\sin^{-1}$.  Our alignment scans show no increase in the integrated intensity of the $(1,1,1)$ Bragg peak between $\mu_{0}H=0$ and $1.37$~T which would indicate a field-induced ferromagnetic component of $\bm{\mu}$ along its direction.  However, a small value for $\mu$ would make it challenging to detect the weak magnetic signal on top of a structural Bragg peak.

Lastly, we comment on the persistence of the broad component of the $(\frac{1}{2},\frac{1}{2},\frac{3}{2})$ magnetic diffraction peak with increasing field.  A possibility is that the component corresponds to magnetic quasielastic or inelastic scattering arising from longitudinal magnetic fluctuations rather than static magnetism. The energy scale associated with such fluctuations would need to be very small, i.e.\ within the $\Delta E\approx0.09$ and $0.07$~meV energy resolutions of the SPINS and FLEXX instruments, respectively, which are quite good for triple-axis neutron spectrometers.  A problem with this interpretation is that such scattering would also be expected to appear around $(\frac{1}{2},\frac{1}{2},\frac{1}{2})$ once $\bm{\mu}$ reorients away from $\bm{\tau}$.  This is not observed in our data.  Further, one may typically expect a magnetic field to open a gap in the spin excitation spectrum of the AFM order.  A large enough gap would cause some dynamic spectral weight to move out of the diffraction measurement's energy window, and the intensity of the diffraction peak to consequently decrease with increasing $H$.  Future, highly-specialized inelastic neutron scattering experiments on YbBiPt will be necessary to determine if the diffraction data include a dynamic component or correspond solely to static magnetism.

\section{Conclusion}
We have presented resistivity and neutron diffraction data that illustrate the response of YbBiPt's fragile AFM to applied magnetic fields along $[\bar{1}~1~0]$, $[0~0~1]$, and $[1~1~0]$.  Our results establish that short-range AFM order characterized by $\bm{\tau}=(\frac{1}{2},\frac{1}{2},\frac{1}{2})$ persists across the previously determined AFM boundary and into the FL region despite clear signatures of an AFM transition in data from transport and thermodynamic experiments\cite{Mun_2013}.   The diffraction data for $\mathbf{H}\parallel[\bar{1}~1~0]$ and $T=0.14$~K show that a broad magnetic diffraction peak exists at $(\frac{1}{2},\frac{1}{2},\frac{3}{2})$ which can be fit by a two-Gaussian lineshape consisting of broad- and narrow-Gaussian components, similar to previous results for $\mu_{0}H=0$~T \cite{Ueland_2014}. Both of the peak's components exist for $0\le \mu_{0}H \alt 1.4$~T, and the peak's total intensity reaches a maximum at $\mu_{0}H=0.55(1)$~T, which is near the FL boundary. The magnetic correlation lengths associated with its components more than double between $\mu_{0}H=0$ and $0.6$~T, reaching maximum values of $\xi_{\text{N}}=136(2)$~\AA\ and $\xi_{\text{B}}=39.7(9)$~\AA.  Thus, the intensity of the magnetic diffraction peak and the magnetic-correlation lengths associated with the short-range AFM  are not maximized within the AFM region, but are largest near the nFL-FL boundary, well away from the AFM-nFL boundary determined via resistivity.  The facts that the domain sizes are maximized and that moment reorientation begins, at least for $T=0.14$~K, near this boundary is intriguing, as it may signal a change to the magnetic exchange, or anisotropy, or both concurrent with a change in the fermiology.

Using data for $\mathbf{H}$ applied within the scattering plane along either $[1~1~0]$ or $[0~0~1]$, and assuming the collinear AFM structure reported in Ref.~\onlinecite{Ueland_2014}, we have further shown that field-induced changes to the $(\frac{1}{2},\frac{1}{2},\frac{3}{2})$ diffraction peak's intensity  are consistent with magnetic domains changing size throughout the AFM and nFL regions.  This agrees with the growth of $\xi_{\text{N}}$ and $\xi_{\text{B}}$ with increasing $\mathbf{H}\parallel[\bar{1}~1~0]$ for these regions.  We have also shown that a magnetic diffraction peak at $(\frac{1}{2},\frac{1}{2},\frac{1}{2})$ appears in the nFL region between $\mu_{0}H=0.2$ to $0.4$~T for $\mathbf{H}\parallel[\bar{1}~1~0]$, and that its appearance signals a reorientation of $\bm{\mu}$ away from $\bm{\tau}$ for $0.4\alt\mu_{0}H\alt1.2$~T.

For $\mathbf{H}\parallel[\bar{1}~1~0]$, the lineshapes of the narrow component of the $(\frac{1}{2},\frac{1}{2},\frac{3}{2})$ peak and the $(\frac{1}{2},\frac{1}{2},\frac{1}{2})$ peak follow a similar field dependence and disappear at $\mu_{0}H\approx1.4$~T.   On the other hand, the broad component of the $(\frac{1}{2},\frac{1}{2},\frac{3}{2})$ peak exists up to at least $\mu_{0}H=1.6$~T, and we discussed that its persistence in field may mean that it is associated with very low energy AFM spin fluctuations rather than static order.  The broadness of the magnetic diffraction peaks, their apparent two-component nature, and their survival across the AFM boundary determined via detailed thermodynamic and transport experiments present an interesting quandary that warrants future investigations into YbBiPt's fragile magnetic properties.

\begin{acknowledgments}
Work at the Ames Laboratory was supported by the Department of Energy, Basic Energy Sciences, Division of Materials Sciences \& Engineering, under Contract No.~\mbox{DE-AC$02$-$07$CH$11358$}.  S.\ M.\ Saunders is supported by the Gordon and Betty Moore Foundation EPiQS Initiative (Grant No.\ GBMF$4411$).  The work at Simon Fraser University was supported by Natural Sciences and Engineering Research Council of Canada.  We acknowledge the support of the National Institute of Standards and Technology, U.\ S.\ Department of Commerce in providing facilities used in this work.   The identification of any commercial product or trade name does not imply endorsement or recommendation by the National Institute of Standards and Technology.  Some measurements were carried out at the E-$4$ and FLEXX instruments at Helmholtz-Zentrum Berlin f\"{u}r Materialien und Energie.  
\end{acknowledgments}


\begin{thebibliography}{99}
\bibitem{Canfield_2016} P. C. Canfield and S. L. Bud'ko, Rep. Prog. Phys. \textbf{79}, 084506 (2016).

\bibitem{Si_2010} Q. Si and F. Steglich, Science \textbf{329}, 1161 (2010).

\bibitem{Stockert_2011} O. Stockert and F. Steglich, Annu. Rev. Condens. Matter Phys. \textbf{2}, 79 (2011).

\bibitem{Stewart_2001} G. R. Stewart, Rev. Mod. Phys. \textbf{73}, 797 (2001).

\bibitem{Hertz_1976}  J. A. Hertz, Phys. Rev. B \textbf{14}, 1165 (1976).

\bibitem{Millis_1993} A. J. Millis, Phys. Rev. B \textbf{48}, 7183 (1993).
	
\bibitem{Mun_2013} E. D. Mun, S. L. Bud'ko, C. Martin, H. Kim, M. A. Tanatar, J.-H. Park, T. Murphy, G. M. Schmiedeshoff, N. Dilley, R. Prozorov, and P. C. Canfield, Phys. Rev. B \textbf{87}, 075120 (2013).

\bibitem{Canfield_1991} P. C. Canfield, J. D. Thompson, W. P. Beyermann, A. Lacerda, M. F. Hundley, E. Peterson, Z. Fisk, and H. R. Ott, J. Appl. Phys. \textbf{70}, 5800 (1991).

\bibitem{Fisk_1991} Z. Fisk, P. C. Canfield, W. P. Beyermann, J. D. Thompson, M. F. Hundley, H. R. Ott, E. Felder, M. B. Maple, M. A. Lopez de la Torre, P. Visani, and C. L. Seaman, Phys. Rev. Lett. \textbf{67}, 3310 (1991).

\bibitem{Robinson_1995} R. A. Robinson, M. Kohgi, T. Osakabe, F. Trouw, J. W.  Lynn, P. C. Canfield, J. D. Thompson, Z. Fisk, and W. P. Beyermann, Phys. Rev. Lett. \textbf{75}, 1194 (1995).

\bibitem{Ueland_2015} B. G. Ueland, S. M. Saunders, S. L. Bud'ko, G. M. Schmiedeshoff, P. C. Canfield, A. Kreyssig, and A. I. Goldman, Phys. Rev. B \textbf{92}, 184111 (2015).

\bibitem{Hundley_1997} M. F. Hundley, J. D. Thompson, P. C. Canfield, and Z. Fisk, Phys. Rev. B \textbf{56}, 8098 (1997).

\bibitem{Movshovich_1994} R. Movshovich, A. Lacerda, P. C. Canfield, J. D. Thompson, and Z. Fisk, Phys. Rev. Lett. \textbf{73}, 492 (1994).

\bibitem{Lacerda_1993} A. Lacerda, R. Movshovich, M. F. Hundley, P. C. Canfield, D. Arms, G. Sparn, J. D. Thompson, Z. Fisk, R. A. Fisher, N. E. Phillips, and H. R. Ott, J. Appl. Phys. \textbf{73}, 5415 (1993).

\bibitem{Movshovich_1994b} R. Movshovich, A. Lacerda, P. C. Canfield, J. D. Thompson, and Z. Fisk, J. Appl. Phys. \textbf{76} 6121 (1994).

\bibitem{Canfield_1994} P. C. Canfield, R. Movshovich, R. A. Robinson, J. D. Thompson, Z. Fisk, W. P. Beyermann, A. Lacerda, M. F. Hundley, R. H. Heffner, D. E. MacLaughlin, F. Trouw, H. R. Ott, Physica B \textbf{197}, 101 (1994).

\bibitem{Ueland_2014} B. G. Ueland, A. Kreyssig, K. Proke\v{s}, J. W. Lynn, L. W. Harriger, D. K. Pratt, D. K. Singh, T. W. Heitmann, S. Sauerbrei, S. M. Saunders, E. D. Mun, S. L. Bud'ko, R. J. McQueeney, P. C. Canfield, and A. I. Goldman, Phys. Rev. B  \textbf{89} 180403(R) (2014).

\bibitem{Robinson_1994} R. A. Robinson, A. Purwanto, M. Kohgi, P. C. Canfield, T. Kamiyama, T. Ishigaki, J. W. Lynn, R. Erwin, E. Peterson, and R. Movshovich, Phys. Rev. B \textbf{50}, 9595 (1994).

\bibitem{Amato_1992} A. Amato, P. C. Canfield, R. Feyerherm, Z. Fisk, F. N. Gygax, R. H. Heffner, D. E. MacLaughlin, H. R. Ott, A. Schenck, and J. D. Thompson, Phys. Rev. B \textbf{46}, 3151 (1992).

\bibitem{Mun_2015} E. D. Mun, S. L. Bud'ko, Y. Lee, C. Martin, M. A. Tanatar, R. Prozorov, and P. C. Canfield, Phys.Rev. B \textbf{92}, 085135 (2015).

\bibitem{Canfield_1992} P. C. Canfield and Z. Fisk, Philos. Mag. B \textbf{65}, 1117 (1992).

\bibitem{Prokes_2017} K. Proke\v{s} and F. Yokaichiya, JLSRF, \textbf{3}, A104 (2017).

\bibitem{Le_2013} M. D. Le, D. L. Quintero-Castro, R. Toft-Petersen, F. Groitl, M. Skoulatos, K. C. Rule, and K. Habicht, Nucl. Instr. and Meth. A \textbf{729}, 220 (2013).

\bibitem{DAVE} R. T. Azuah, L. R. Kneller, Y. Qiu, P. L. W. Tregenna-Piggott, C. M. Brown, J. R. D. Copley, and R. M. Dimeo, J. Res. Natl. Inst. Stan. Technol. \textbf{114}, 341 (2009).

\bibitem{LAMP} \url{https://www.ill.eu/users/support-labs-infrastructure/software-scientific-tools/lamp/download-links/}

\bibitem{SPECTRA} \url{https://github.com/simonward86}

\bibitem{Lynn_1994} J. W. Lynn, J. Appl. Phys. \textbf{75}, 6806 (1994).

\bibitem{Fawcett_1988} E. Fawcett, Rev. Mod. Phys. \textbf{60}, 209 (1988).

\bibitem{Werner_1967} S. A. Warner, A. Arrott, and H. Kendrick, Phys. Rev. \textbf{155}, 528 (1967).

\bibitem{Nakotte_1993} H. Nakotte, R. A. Robinson, J. W. Lynn, E. Br\"{u}ck, and F. R. de Boer, Phys. Rev. B \textbf{47}, 831 (1993).

\bibitem{Nakotte_1998} H. Nakotte, R. A. Robinson, A. Purwanto, Z. Tun, K. Proke\v{s}, E. Br\"{u}ck, and F. R. de Boer, Phys. Rev. B \textbf{58}, 9269 (1998).

\bibitem{Stockert_2007} O. Stockert, M. Enderle, and H. v. L\"{o}hneysen, Phys. Rev. Lett. \textbf{99}, 237203 (2007).

\bibitem{Lohneysen_2007} H. v. L\"{o}hneysen,A. Rosch, M. Vojta, and P. W\"{o}lfle, Rev. Mod. Phys. \textbf{79}, 1015 (2007).

\bibitem{Shirane_2002} G. Shirane, S. M. Shapiro, and J. M. Tranquada, \textit{Neutron Scattering with a Triple-Axis Spectrometer} (Cambridge University Press, Cambridge, 2002), pp. 36--46.

\bibitem{McIntyre_1988} G. J. McIntyre and R. F. D. Stansfield, Acta. Cryst. A\textbf{44}, 257 (1988).

\bibitem{Axe_1983} J. D. Axe and J. B. Hastings, Acta. Cryst. A\textbf{39}, 593 (1983).

\end{thebibliography}
\end{document}